\documentclass{aa}
\usepackage{graphicx}
\usepackage{rotating}
\usepackage{placeins}

\def\rosat{{\sl ROSAT }}

\def\chandra{{\sl Chandra }}
\def\xmm{{\sl XMM-Newton }}
\def\ergsec{\hbox{erg s$^{-1}$ }}
\def\ergcm{\hbox{erg cm$^{-2}$ s$^{-1}$ }}

\usepackage{color}

\begin{document}

\title{Hot gas and magnetic arms of \object{NGC\,6946}: indications for reconnection heating?
\thanks{
Based on observations obtained with {\sl XMM-Newton}, an ESA science mission with instruments and contributions
directly funded by ESA Member States and NASA
}}
\author{M. We\.zgowiec\inst{1,2}
\and M. Ehle\inst{3}
\and R. Beck\inst{4}}
\institute{Astronomisches Institut der Ruhr-Universit\"at Bochum, Universit\"atsstrasse 150, 44780 Bochum, Germany
\and Obserwatorium Astronomiczne Uniwersytetu Jagiello\'nskiego, ul. Orla 171, 30-244 Krak\'ow, Poland, 
\email{markmet@oa.uj.edu.pl}
\and ESA-ESAC, XMM-Newton Science Operations Centre, P.O. Box 78, 28691 Villanueva de la Ca\~nada, Madrid, Spain
\and Max-Planck-Institut f\"ur Radioastronomie, Auf dem H\"ugel 69, 53121 Bonn, Germany}
\offprints{M. We\.zgowiec}
\date{Received date/Accepted date}

\titlerunning{Hot gas and magnetic arms of...}
\authorrunning{M. We\.zgowiec et al.}

\abstract
{The grand-design face-on spiral galaxy NGC\,6946 is remarkable because of its
high star formation activity, the massive northern spiral arm, and the
magnetic arms, which are observed in polarized radio synchrotron emission and are
located between the optical arms and possibly are magnetic reconnection regions. 
}
{
We used electron densities and temperatures in star-forming (active) and less active regions
and compared them to findings from the analysis
of the radio data to study the energy budget of NGC\,6946. 
The hot gas
above the magnetic arms between the optical arms might suggest
gas heating by reconnection.
We also study the population of point sources in NGC\,6946, including the origin
of the puzzling ultra-luminous emission complex MF16.
}
{
X-ray observations of NGC\,6946 performed with \xmm were used to
study the emission from X-ray point sources and
diffuse hot gas, including the magnetic arms and the halo.
Spectral fitting of the diffuse X-ray emission allowed us to derive temperatures
of the hot gas. With assumptions about the emission volume, this allowed us to estimate gas
densities, masses, and cooling times.
}
{
To explain the X-ray emission from the spiral arms of NGC\,6946 two-temperature plasma
models are needed to account for the disk and halo emission. The interarm regions show
only one thermal component. We observe that the temperature of the hot gas
in and above the magnetic arm regions increases slightly when compared to the average temperatures in the 
areas in and above the spiral arms. 
For the southwestern part of the disk, which is depolarized in the radio range by Faraday rotation, 
we find more efficient mixing of disk and halo gas. 
}
{
We propose magnetic reconnection in the magnetic arm regions of NGC\,6946 as the possible 
cause of the additional heating of the gas and ordering of the magnetic fields. 
In the southwestern part of the galactic disk 
we observed indications of a possible faster outflow of the hot gas. 
A very hot gas within the MF\,16 nebula possibly suggests shock heating by a supernova explosion.
}

\maketitle

\section{Introduction}

NGC\,6946 (see Table~\ref{astrdat}) is a Scd spiral galaxy seen
face-on, which is listed in Arp's
atlas of peculiar galaxies because of its massive northern spiral
arm. Its large optical diameter and low inclination make NGC\,6946 one of
the most prominent grand-design spiral galaxies in the sky. It is known
to host a bright starburst nucleus \citep[see e.g.][]{telesco}.
Detections of CO radio emission lines \citep{nieten,walsh} gave further evidence for
the high star-forming activity of NGC\,6946 (especially in the northern spiral arm),
proved that molecular clouds containing warm and dense gas are distributed
throughout the inner disk, and showed that the total molecular gas mass
is exceptionally high for a spiral galaxy. The distributions of the
emission from NGC\,6946 in various spectral ranges were analysed with wavelet functions \citep[see][]{frick}.

\begin{table}[ht]
        \caption{\label{astrdat}Basic astronomical properties of NGC\,6946}
\centering
                \begin{tabular}{lccccc}
\hline\hline
Morphological type & SABc       \\
Inclination        & 18\degr    \\
Diameter D$_{25}$  & 11\farcm5  \\
R.A.$_{2000}$      & 20$^{\rm h}$34$^{\rm m}$53$^{\rm s}$       \\
Dec$_{2000}$       & +60\degr 09\arcmin 13\arcsec       \\
Distance\tablefootmark{a} & 7\,Mpc      \\
\hline
\end{tabular}
\tablefoot{
All data except for the distance are taken from the HYPERLEDA database -- http://leda.univ-lyon1.fr -- see \citet{leda}.\\
\tablefoottext{a}{Taken from \citet{israel80}.}
}
\end{table}

\subsection{Previous radio continuum findings}
NGC\,6946 has been thoroughly studied in the radio regime in total and
polarized emission \citep{beck91,ehleN6946}.
Radio polarization observations at 18\,cm and 20\,cm wavelengths
 revealed excess Faraday rotation and
strong depolarization in the SW quadrant of NGC\,6946 that is probably due to a large-scale magnetic field along
the line of sight, oriented approximately perpendicular to the disk plane \citep{beck91}.
Analysis of polarization data at four frequencies also
suggested strong vertical fields extending far above the disk \citep{beck07}.
Such field lines should enable an outflow of hot gas into a halo.
As the SW quadrant of NGC\,6946 is a region of relatively low star-formation
activity, this phenomenon resembles a {\it \textup{coronal hole}} on the Sun.

The average energy density of the warm ionized medium in the interstellar medium (ISM) of the inner
disk of NGC\,6946 was found to be lower by a factor of about 10 than the
energy densities of the magnetic field and that of the cosmic rays, resulting
in the conclusion that the magnetic field dominates thermal processes in the
disk and halos of galaxies \citep{beck04}. A significant fraction of the
diffuse ISM must be unstable, giving rise to gas outflows.

Radio polarization data at $\lambda6$~cm led to the discovery of a new
phenomenon: highly aligned magnetic fields that are concentrated in two
main spiral features, located almost precisely {\it \textup{between}} the optical
spiral arms of NGC\,6946 \citep[][their Fig.~1]{beckhoernes}.
No enhanced densities of molecular, neutral, or warm ionized gas have been
detected at the positions of these magnetic arms. However, radio
observations revealed significant Faraday rotation in these regions so that
some ionized gas must be present.
\citet{frick} analysed the magnetic and optical spiral arms in
NGC\,6946 using 1D wavelet transformations and found that each magnetic arm is similar
to the {\sl \textup{preceding}} optical arm and hence can be regarded as a
phase-shifted image.

Rudimentary magnetic arms were also found in other spiral galaxies \citep{beck15},
but NGC\,6946 still is the most prominent example.

The magnetic arms contradict density-wave models, which predict
enhanced ordered magnetic fields at the inner edges of the arms.
Several mechanisms were proposed to explain the magnetic arms.
For example, the continuous injection and amplification of turbulent fields
by supernova shock fronts may suppress the mean-field dynamo in the material arms \citep{moss13,moss15}.
Alternatively, the introduction of a relaxation time of the magnetic response in the dynamo equation may
lead to a phase shift between the material and magnetic spiral arms \citep{chamandy13a,chamandy13b}.
The mean-field dynamo in the material arms might also be suppressed by outflows driven by star
formation \citep{chamandy15}.

We propose here that the strongly polarized radio emission from the magnetic arms may
also suggest the existence of reconnection regions where cosmic rays are accelerated.
The gas heated by the same process should be detectable in the X-ray domain \citep{lesch,hanasz}.
By comparing the properties of the hot gas in the magnetic and spiral arms, we may be able 
to trace a possible additional heating of the gas that would be caused by the reconnection processes.

\subsection{Earlier X-ray observations}
Despite the rather high $N_{\rm H}$-value of about $\rm 2\times10^{21}$~cm$^{-2}$ (see Tab.~\ref{obsred}),
NGC\,6946 was detected by {\sl ROSAT}'s All Sky survey with a
count rate of 0.1~cts~s$^{-1}$. Analysing the 36 ks \rosat PSPC
pointed observation, \citet{schlegel94a, schlegel94b, schlegel94c}
apart from SN 1980K additionally reported emission from nine point-like
X-ray sources and diffuse emission from NGC~6946. The brightest source
(identified at that time with a very luminous supernova remnant MF16) has a count rate of
0.07~cts~s$^{-1}$ corresponding to a flux of $8.3\times 10^{-13}$~\ergcm
in the energy range 0.5-2 keV, three sources are fainter by about a factor of
10, and the rest (fainter by a factor of 40) are at the
detection threshold.

\citet{holt} studied discrete X-ray sources in NGC\,6946
using a 60 ks \chandra ACIS observation and found the source population
dominated by high-mass X-ray binaries. Their survey was
complete down to approximately $10^{37}$~\ergsec. However, in contrast to
previous results, the ultra-luminous MF16 complex was found to be
deficient in line emission expected from an interaction with a dense
surrounding medium. Its spectrum lacks pronounced spectral lines and can be
fit with a variety of models that are all associated with unusually high luminosities,
leaving the origin of the MF16 related X-ray emission unknown.
\citet{schlegel03} used the same \chandra observations and the
luminosity function derived by \citet{holt} to distinguish point
sources
from the diffuse emission. These authors estimated that as much as 10\% of the
total soft X-ray emission could be due to a hot diffuse component.

We checked the \chandra archive and found additional
ACIS-S pointings: one on-axis pointing aiming at SN2002HH (30~ks, IAU Circ. 8024
and Roberts \& Colbert 2003), and three ($30$~ks each) off-axis pointings centred
on SN2004et. The combination of these \chandra observations was used
by severeal authors \citep{soria08, fridriksson08, kajava09, liu11}
to investigate individual sources (e.g. supernovae, ultra-luminous sources) in NGC\,6946
and their spectral and temporal variability.
Recent observations (20~ks, PI Kochanek) were also aimed at studying point-source populations, which is beyond the scope of this paper.

\subsection{Immediate objectives}

This paper focuses on a detailed analysis of the extended emission from the hot gas of NGC\,6946 with the use of the data acquired 
by the XMM-Newton X-ray telescope \citep{jansen}. The parameters of the hot gas acquired
from the spectral analysis of selected regions of the galaxy are compared with the properties of the radio emission, especially with
its polarized component that traces the structure of the magnetic field of the galaxy.

In the following section (Sect.~\ref{obsred}) details of the data reduction and analysis are presented. Section~\ref{results} presents the distribution of the X-ray emission from both diffuse gas and point-source populations, and we also describe the spectra
we obtained.
In Sect.~\ref{disc} we thoroughly discuss the results, including correlations and comparisons
with the polarized radio emission from NGC\,6946. We also provide
new insight into the nature of the ultra-luminous source MF16.
We conclude in Sect.~\ref{cons}.

\section{Observations and data reduction}
\label{obsred}

\begin{table}[ht]
        \caption{\label{xdat}Characteristics of the XMM-Newton X-ray observations of
        NGC\,6946}
\centering
                \begin{tabular}{lccccc}
\hline\hline
Obs ID                            & 0200670301  \\
                                  & 0500730101  \\
                                  & 0500730201  \\
                                  & 0691570101  \\      
column density $N_{\rm H}$\tablefootmark{a}& 1.84   \\
MOS filter                        & medium   \\
MOS obs. mode                     & FF   \\
pn filter                         & medium   \\
pn obs. mode                      & FF   \\
Total/clean pn time [ks]          & 13.1/8.3   \\
                                  & 28.4/20.2  \\
                                  & 33.3/29.7  \\
                                  & 114.3/98.2 \\
\hline
\end{tabular}
\tablefoot{
\tablefoottext{a}{Column density in [10$^{21}$ cm$^{-2}$] weighted average value after LAB Survey of Galactic \ion{H}{i} 
\citet{lab}.}
}
\end{table}

NGC\,6946 has been observed 11 times between 2003 and 2006 with the XMM-Newton telescope \citep{jansen},
but the observations always suffered from heavy high flaring radiation.
The only observation that yielded any clean data was made on 13 June 2004 (ObsID 0200670301).
Consequently, the galaxy was observed again, on 2 and 8 November 2007 (ObsIDs 0500730201 and 0500730101, respectively,
see Table~\ref{xdat}). The data were still affected by
high flaring radiation, but this time it was possible to obtain many good-quality data.
The effect of frequent flaring radiation on the observations was caused by visibility constraints that required
NGC\,6946 to be observed at the end of an XMM-Newton revolution. In recent
years the orbit has evolved and NGC\,6946 can be
much better observed. Since early 2012 it has become possible to observe this galaxy
for almost a full orbit (144\,ks).
The most recent observations, performed between 21 and 23 of December 2012 
and aimed at the ultra-luminous source NGC\,6946 X-1 (ObsID 0691570101), provided a long exposure that was relatively free of high flaring
radiation; this resulted in 98\,ks of good data.

The data were processed using the SAS 13.0.0 package \citep{sas}
with standard reduction procedures. Following the routine of tasks $epchain$ and $emchain$,
event lists for two EPIC-MOS cameras \citep{turner} and the EPIC-pn camera \citep{strueder}
were obtained. Next, the event lists were carefully filtered for periods of intense radiation
of high-energy background by creating light curves of high-energy emission. These light curves were used to produce
good time interval (GTI) tables, which mark times of low count
rates of high-energy emission. Such tables (time ranges) were then used to remove the remaining data
when high count rates were observed. The resulting lists were checked for the residual existence
of soft proton flare contamination, which
could influence the faint extended emission. To do that, we used a script\footnote{http://xmm2.esac.esa.int/external/xmm\_sw\_cal/\\background/epic\_scripts.shtml\#flare}
that performs calculations developed by \citet{spcheck}.
We found that only the shortest observation (ObsID 0200670301) is contaminated very slightly by soft proton radiation. To ensure the best-quality data (crucial to analyse diffuse emission), we only used events with FLAG=0 and PATTERN$\leq$4 (EPIC-pn) or FLAG=0 and PATTERN$\leq$12 (EPIC-MOS) in the following data
processing.

The filtered event lists were used to produce images, background images, exposure maps (without vignetting correction),
masked for an acceptable detector area using the images script\footnote{http://xmm.esac.esa.int/external/xmm\_science/\\gallery/utils/images.shtml}, modified by the authors
to allow adaptive smoothing.
All images and maps were produced (with exposure correction) in four energy bands of 0.2 - 1 keV, 1 - 2 keV, 2 - 4.5 keV,
and 4.5 - 12 keV.
The images were then combined into final EPIC images and adaptively smoothed with a maximum smoothing scale of 30$\arcsec$ FWHM.
The rms values were obtained by averaging the emission over a large source-free area in the final map.

Another set of images was also constructed after excluding all point sources found within the D$_{25}$ diameter of NGC\,6946 from the event lists (see below for details on point source exclusion).
This was done with the help of a routine used to create re-filled blank sky
background maps - {\it ghostholes\_ind}\footnote{ftp://xmm.esac.esa.int/pub/ccf/constituents/extras/\\background/epic/blank\_sky/scripts}.
In this way we obtained a map of diffuse emission where all regions of excluded point sources are filled with emission close to extracted regions by sampling adjacent events and
randomising spatial coordinates\footnote{http://xmm2.esac.esa.int/external/xmm\_sw\_cal/\\background/blank\_sky.shtml\#BGsoft}. Although this method is used to handle background maps, we
obtained good results when we applied it to real source data.
Section~\ref{diffuse} presents images of soft (0.2-1\,keV) and medium (1-2\,keV) emission, together with a corresponding 
hardness ratio map, defined as $$HR=\frac{{\rm med}-{\rm soft}}{{\rm med}+{\rm soft}},$$ for images with and without detected point sources.

Next, the spectral analysis was performed. To create spectra we only used the event list from the EPIC-pn camera because
it offers the highest sensitivity in the soft energy band. Only the emission above
0.3\,keV was analysed because the internal noise of the pn camera is too high below this
limit\footnote{http://xmm.esac.esa.int/external/xmm\_user\_support/\\documentation/uhb}. Although this is not crucial
when combined with MOS cameras to produce images, it is important
to exclude the softest emission below 0.3\,keV to
obtain reliable good-quality spectra.

Unsmoothed images for all bands were used to search for point sources with the standard SAS
{\it edetect\_chain} procedure. Regions found to include a possible point source were marked. The area was individually chosen
for each source to ensure that we excluded all pixels brighter than the surrounding background.
These areas were then used to construct spectral regions for which spectra were acquired.

The non-default way of excluding the detected point sources helped to keep more diffuse emission in the final spectra. 
However, expecting some contribution from the PSF wings, we added a power-law component to our model fits to account for any residual emission.
 A power-law component was also needed to account for unresolved point sources.
The background spectra were obtained using blank sky event lists \citep[see][]{carter}.
These blank sky event lists were filtered using the same procedures as for the source event lists.
For each spectrum we produced response matrices and effective area files. For the latter, detector maps needed for extended
emission analysis were also created. The spectra were binned, which resulted in a better signal-to-noise ratio. To obtain a reasonable number of
bins at the same time, we chose to have 25 total counts per energy bin. The spectra were fitted using XSPEC~11 \citep{xspec}.

Since observations 0500730201 and 0500730101 have identical pointings
and position angles, we merged the cleaned event lists using the SAS task {\it merge}. The spectra extracted from this merged lists give the same model-fitting results as separate
spectra fitted simultaneously,  but the errors are better constrained, therefore we used the former spectra in our final analysis. It was not possible to
also merge-in the shortest observation (ObsID 0200670301) because it has different
parameters (pointing and position angle), and simultaneous fitting with the larger data set showed that both spectra are systematically offset. This resulted in a poorer model fit. We assume that this might arise because the observation was significantly affected by high flaring radiation and because the filtered
''clean'' data still show residual contamination that might influence the spectral fitting.
Although we used this observation for image production, we therefore
excluded the pn data from our spectral
analysis. For the same inconsistency reasons (different pointing and position angle), we used the longest observation 0691570101 separately when performing the spectral analysis.
This approach resulted in two corresponding spectra for each of the studied regions. Each pair of spectra was merged (as well as their corresponding background spectra) 
using the SAS task $epicspeccombine$. Although for multiple spectra a most commonly advised routine is a simultaneous fit rather than a fit to a combined spectrum, 
we note that for spectra with a very different sensitivity (certainly in our case), a combination of spectra leads to a much better handle on the background 
and consequently a better fit. This is because background subtraction only takes place for the merged spectrum after the source and background spectra 
are combined (contrary to a simultaneous fit, where each spectrum is background subtracted before the fit).

For the overlays we also used the XMM-Newton Optical Monitor data acquired during the same observations and produced an image in the UVM2 filter 
using the standard SAS {\it omchain} procedure.

\section{Results}
\label{results}

\subsection{Distribution of the X-ray emission}
\label{dist}

\subsubsection{Diffuse emission}
\label{diffuse}

\begin{figure*}[ht]
                        \resizebox{0.46\hsize}{!}{\includegraphics{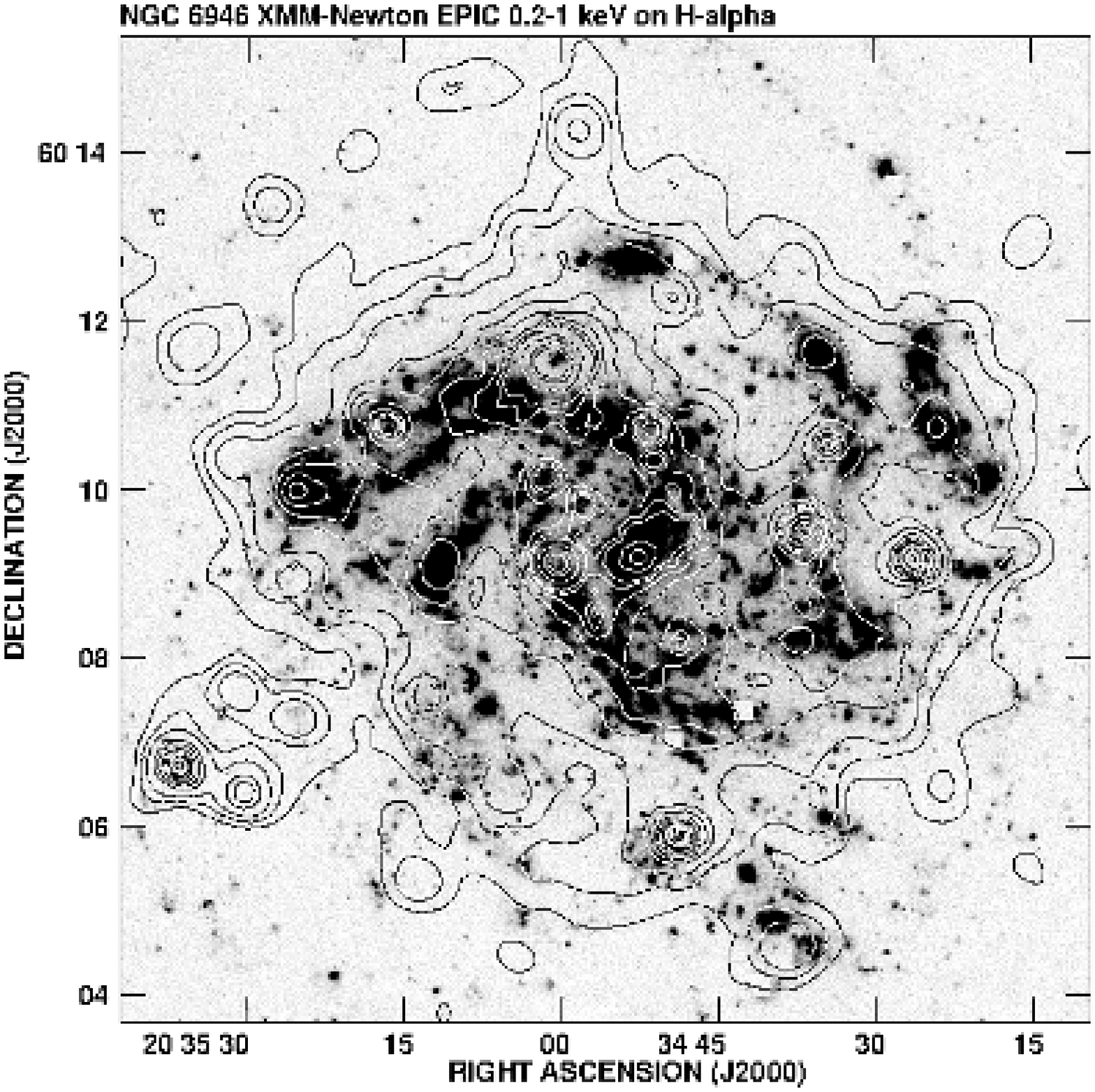}}
                        \resizebox{0.45\hsize}{!}{\includegraphics{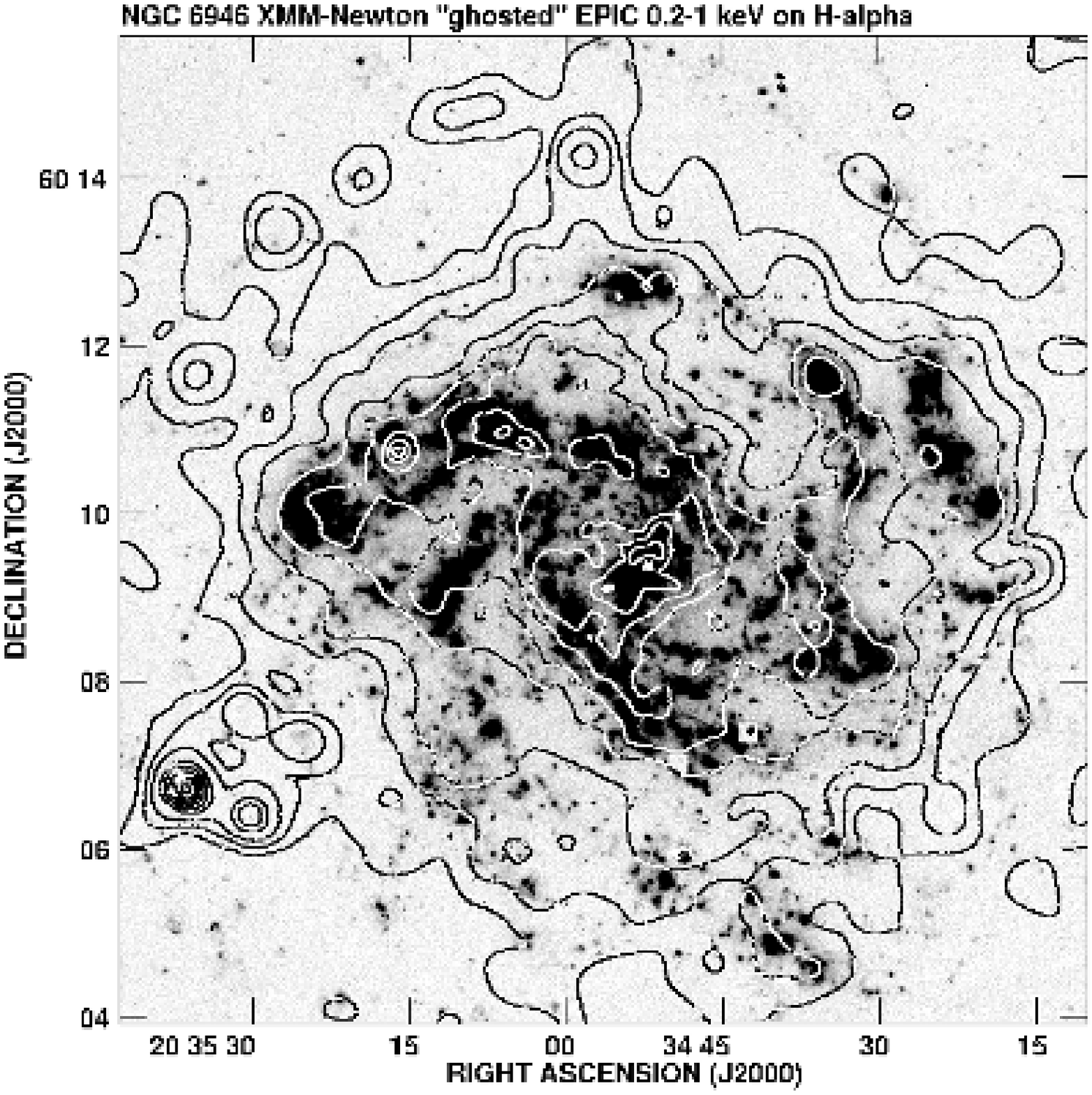}}
                \caption{
                {\it Left:} Map of soft X-ray emission
                from NGC\,6946 in the 0.2 - 1 keV band overlaid onto an H$\alpha$ image. The
                contours are 3, 5, 8, 16, 25, 40, 60, 80, 100, 200, 500, and 1000 $\times$ rms. The map is adaptively smoothed with the
                largest scale of 30$\arcsec$. {\it Right:} Same map, but with point sources excluded from the galactic disk (see text for details).
                }
                \label{6946xsoft}
        \end{figure*}

NGC\,6946 shows soft extended X-ray emission corresponding to the entire star-forming disk (Fig.~\ref{6946xsoft}), with the brightest emission closely following
the star-forming regions. Although the southern part of the star-forming disk is less pronounced than the northern one, no asymmetries of the emission from the hot gas are
visible. In contrast, the X-ray emission seems to extend farther out beyond star-forming regions in the southern part of the galaxy. An area of diffuse emission around
the galactic centre forms a structure resembling a very small bar that crosses the central core and is aligned with the H$\alpha$ emission.

\begin{figure*}[ht]
                        \resizebox{0.46\hsize}{!}{\includegraphics{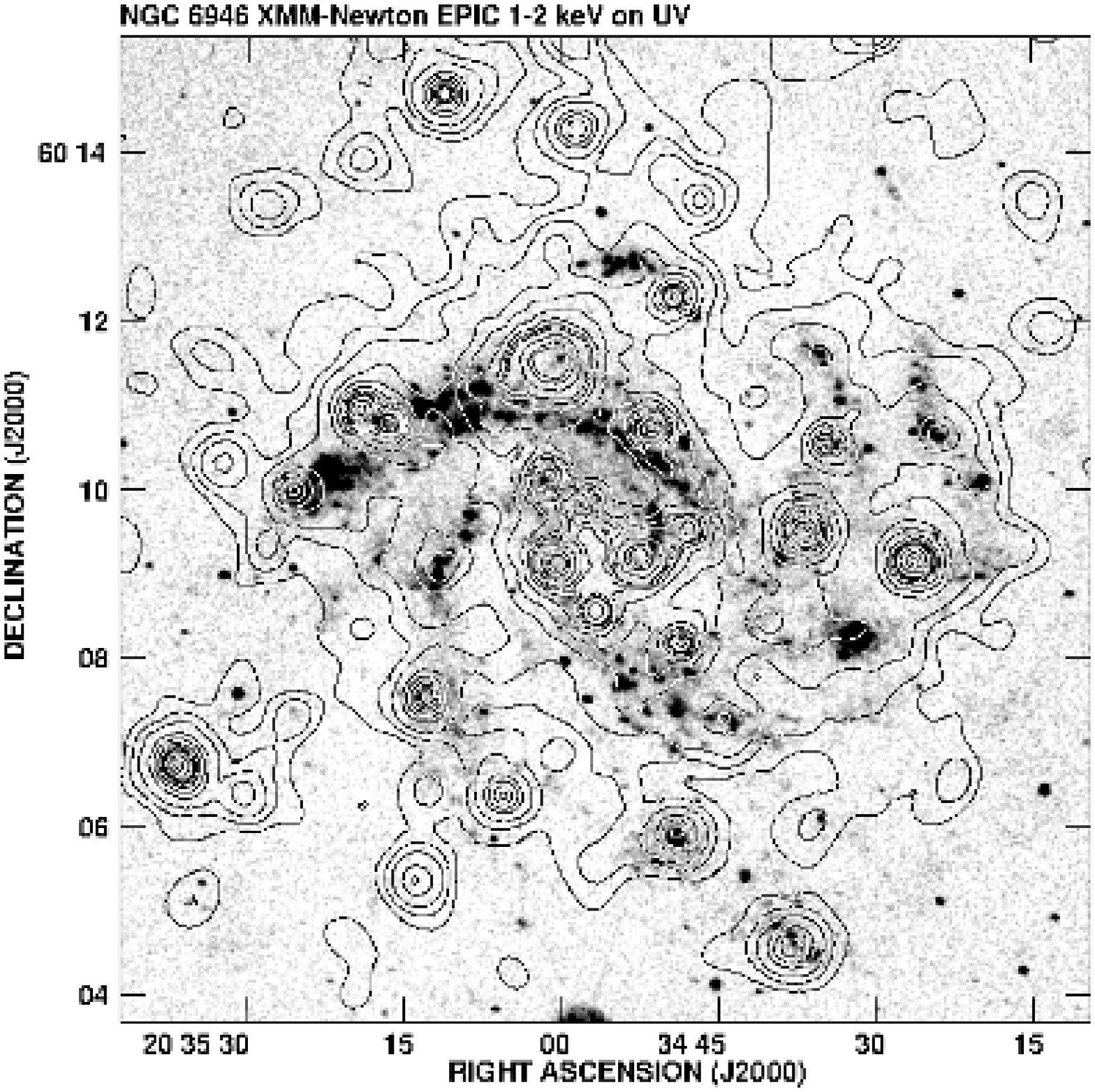}}
                        \resizebox{0.45\hsize}{!}{\includegraphics{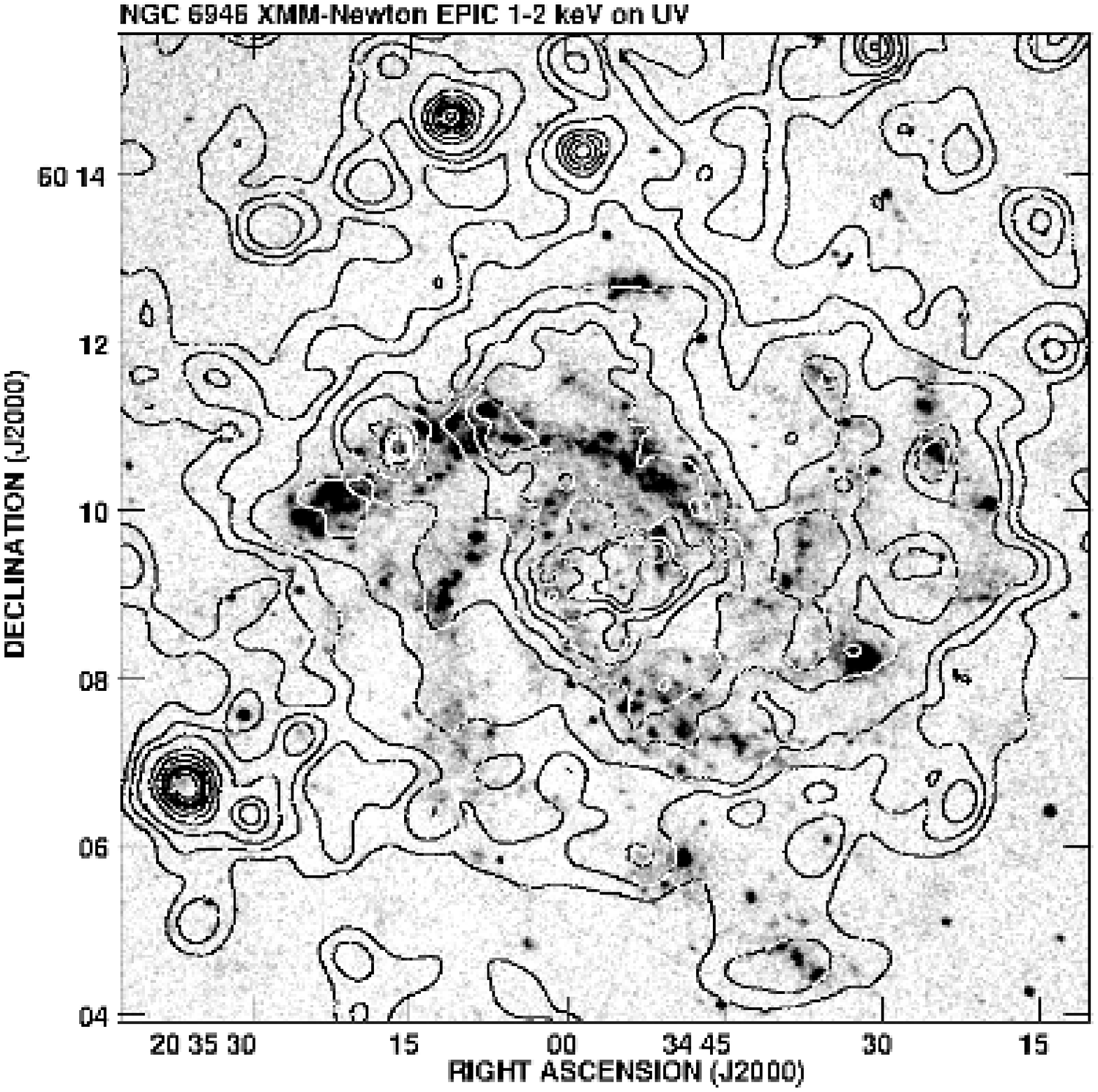}}
                \caption{
                {\it Left:} Map of medium X-ray emission from NGC\,6946 in the 1 - 2 keV band overlaid onto the XMM-Newton Optical Monitor UVM2 filter image.
                The contours are 3, 5, 8, 16, 25, 40, 60, 80, 100, 200, 300, 500, and 1000 $\times$ rms. The map is adaptively smoothed
                with the largest scale of 30$\arcsec$. {\it Right:} Same map, but with point sources excluded from the galactic disk (see text for details).
                }
                \label{6946xmedium}
        \end{figure*}

\begin{figure*}[ht]
                        \resizebox{0.45\hsize}{!}{\includegraphics[clip]{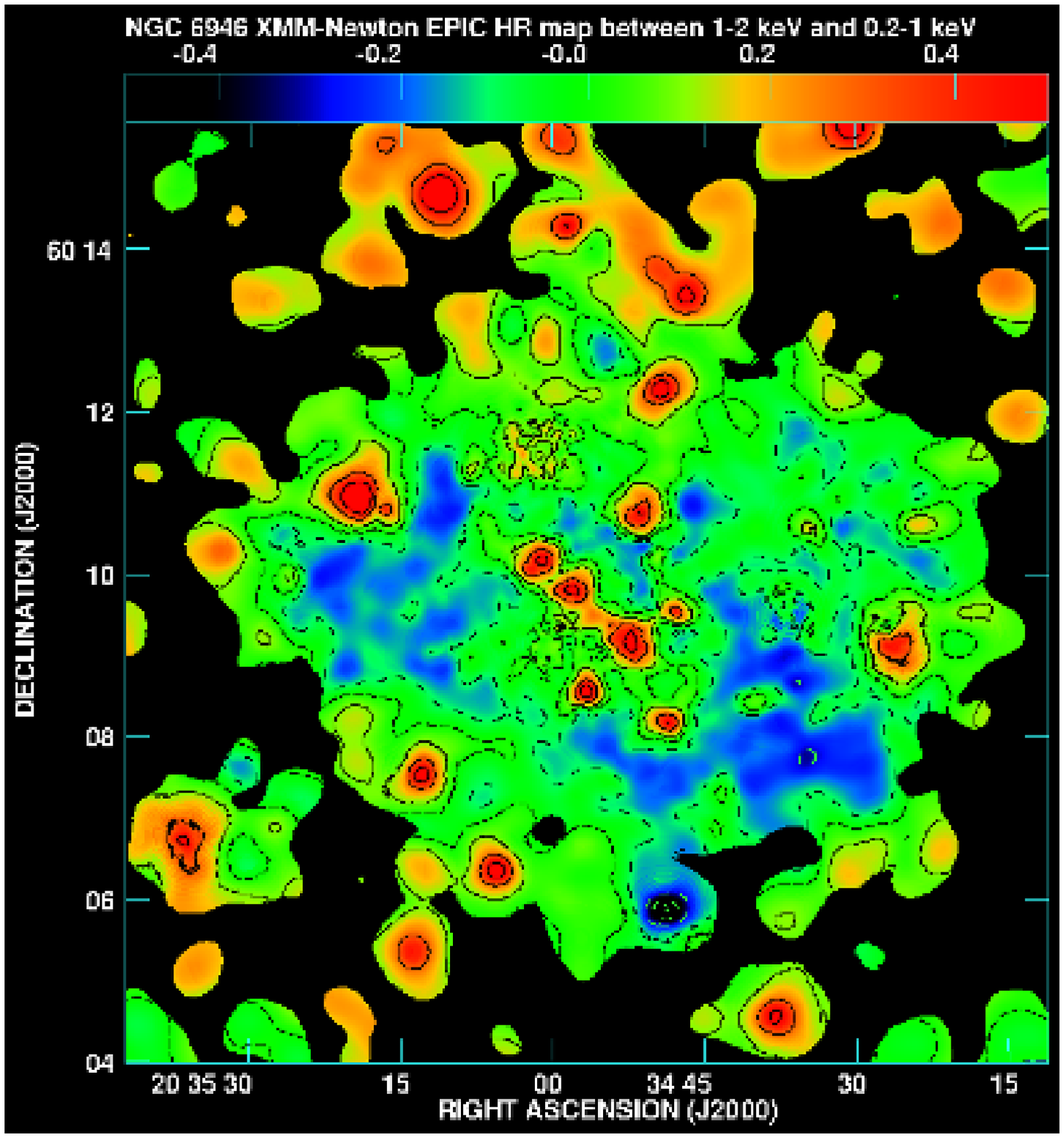}}
                        \resizebox{0.46\hsize}{!}{\includegraphics[clip]{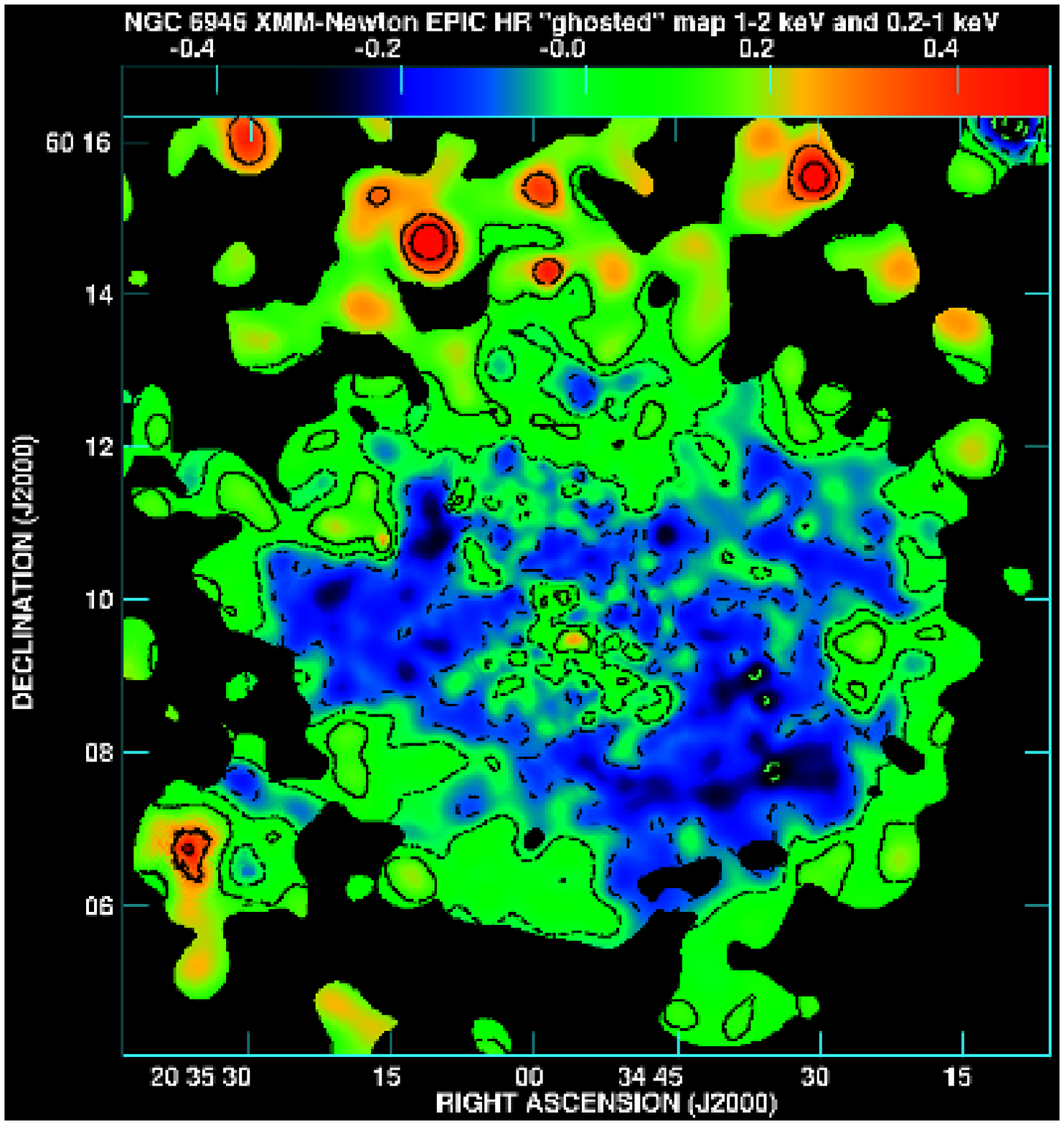}}
                \caption{Maps of the hardness ratio between medium and soft X-ray emission from NGC\,6946 (Figs.~\ref{6946xmedium}
                         and~\ref{6946xsoft}). The map is truncated at the 3$\sigma$ level of the 1-2\,keV map. {\it Left:} Map with point sources. 
                        {\it Right:} Map without point sources.}
                \label{6946hr}
        \end{figure*}

The hot gas disk visible in the 1-2\,keV energy band (Fig.~\ref{6946xmedium}) is extended in a similar way as the emission in the softer energy band (Fig.~\ref{6946xsoft}).
This may suggest large amounts of very hot gas in the galactic disk and/or halo. To further investigate the contribution from the
hottest gas to the X-ray emission from NGC\,6946, we produced a hardness ratio map using both distributions (Fig.~\ref{6946hr}).

As mentioned before, the two distribution are similar on average because the values in most parts of the HR map are close to 0. Nevertheless, north-east and south-west of the centre,
distinct areas of softer emission are clearly visible.
This corresponds well to the orientation of the bright star-forming regions visible in the H$\alpha$ map (Fig.~\ref{6946xsoft}).
However, the softest emission is produced in the south-western part of the disk, 
where the production of young massive stars is diminished, as seen in the UV map (Fig.~\ref{6946xmedium}).

\subsubsection{Point sources}
\label{pointdist}

\begin{figure*}[ht]
\begin{center}
\resizebox{0.46\hsize}{!}{\includegraphics[clip]{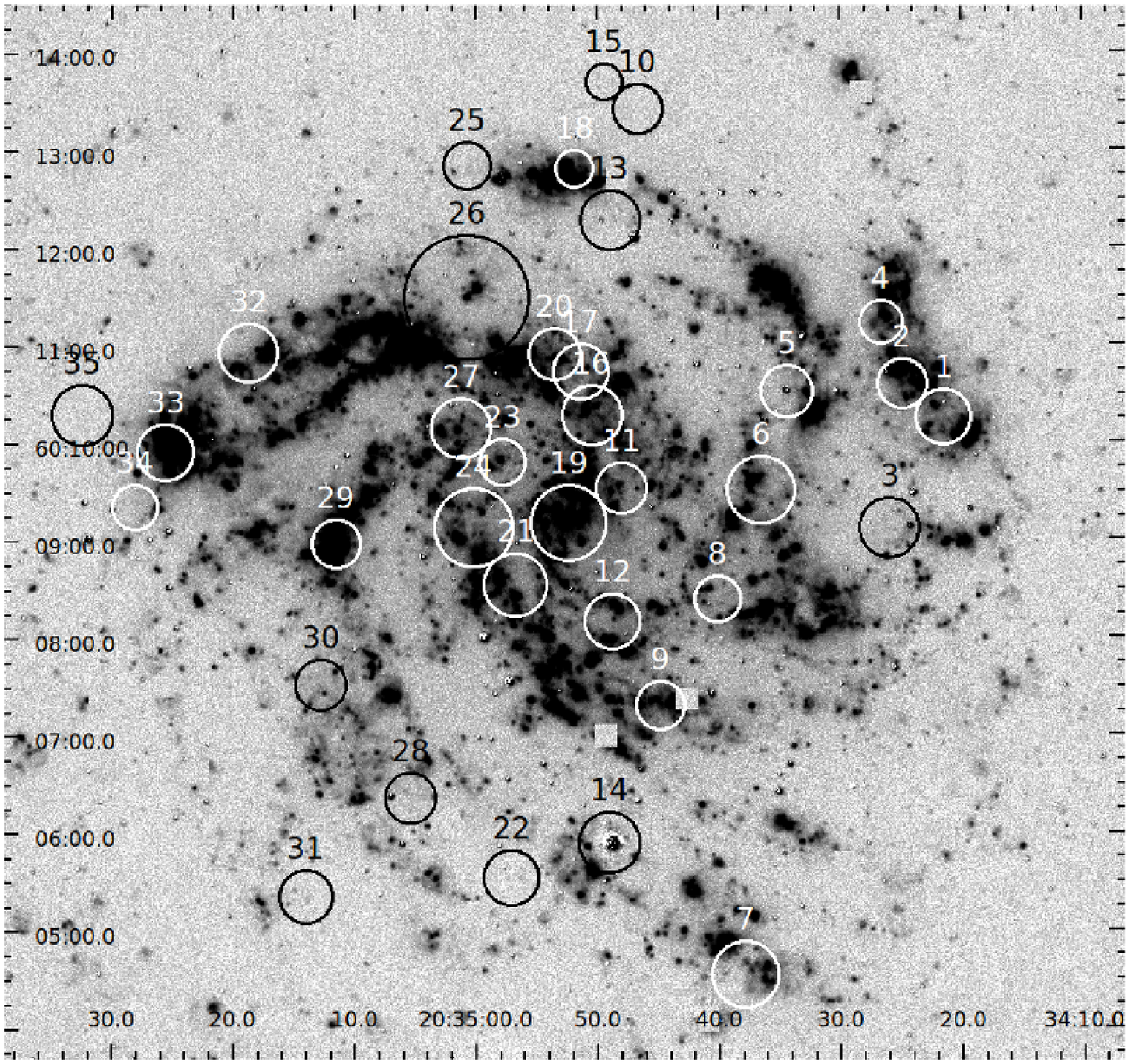}}
\resizebox{0.45\hsize}{!}{\includegraphics{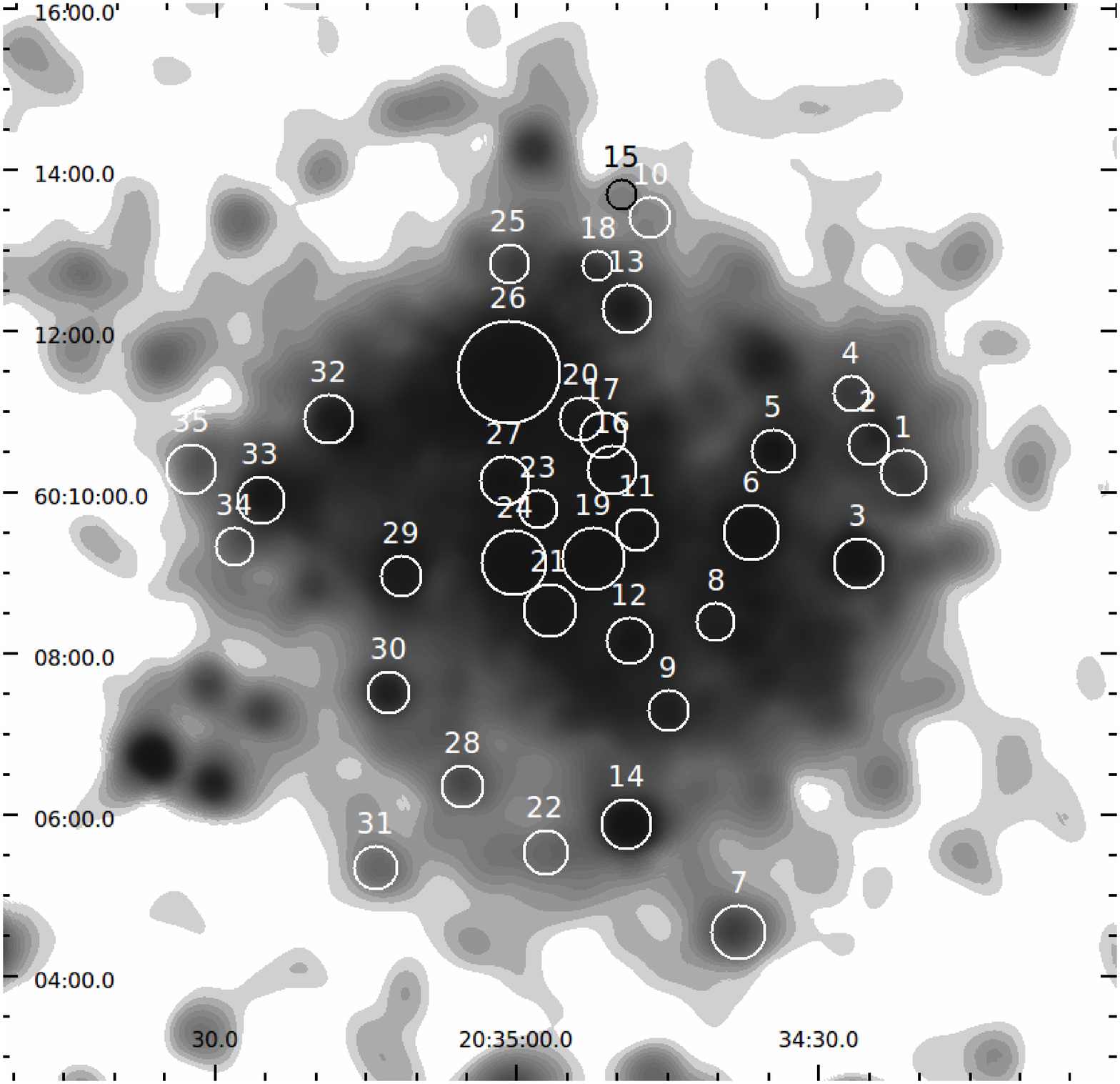}}
\end{center}
\caption{
         {\it Left}: All point-like sources detected in NGC\,6946 (see text for a detailed description) overlaid on an H$\alpha$ image.
         {\it Right}: The same regions as in the left panel overlaid on the map of soft X-ray emission in the 0.2 - 1 keV band
                     as shown in Fig.~\ref{6946xsoft}.
        }
                \label{6946points}
        \end{figure*}

Figure~\ref{6946points} shows all detected point sources within the D$_{25}$ disk of NGC\,6946. 
For all sources we performed a spectral analysis.
For weak sources the hardness ratios (HRs) were derived (see Table~\ref{6946sources}).
We used the same energy bands as in \citet{pietsch04}: (0.2-0.5)\,keV, (0.5-1.0)\,keV, (1.0-2.0)\,keV, (2.0-4.5)\,keV,
and (4.5-12)\,keV as bands 1 to 5. Consequently, the hardness ratios
are calculated as $HR_i=B_{i+1}-B_i/B_{i+1}+B_i$ for $i$ = 1 to 4, where $B_i$ is the count rate in band $i$, as defined above.
For sources with more than 500 net counts
in the total energy band (0.2-12\,keV), spectra were extracted and fitted with models (see Sect.~\ref{sourcespec}).

\begin{sidewaystable*}[ht]
\caption{\label{6946sources}Countrates and hardness ratios for the resolved point-like sources in NGC\,6946}
\centering
\begin{tabular}{rllrrrrrrrrrrr}
ID & $\textstyle\alpha_{2000}$ & $\textstyle\delta_{2000}$ & R & B$_1$ & B$_2$ & B$_3$ & B$_4$ & B$_5$ & C$_t$ & HR$_1$ & HR$_2$ & HR$_3$ & HR$_4$ \\
(1) & (2) & (3) & (4) & (5) & (6) & (7) & (8) & (9) & (10) & (11) & (12) & (13) & (14) \\
\hline\hline
1 &20$^{\rm h}$34$^{\rm m}$21\fs4 & +60\degr 10\arcmin 15\farcs2& 17 & 5 & 55 & 17 & 34 & 29 & 140 & 0.83 & -0.53 & 0.33 & -0.08 \\
2 &20$^{\rm h}$34$^{\rm m}$24\fs9 & +60\degr 10\arcmin 35\farcs6& 15 & 7 & 87 & 102 & 87 & 86 & 369 & 0.85 & 0.08 & -0.08 & -0.01 \\
{\bf 3} &{\bf 20$^{\rm h}$34$^{\rm m}$25\fs9} & {\bf +60\degr 09\arcmin 07\farcs0}& {\bf 19} & {\bf 52} & {\bf 349} & {\bf 577} & {\bf 322} & {\bf 68} & {\bf 1368} & {\bf 0.74} & {\bf 0.25} & {\bf -0.28} & {\bf -0.65} \\
4 &20$^{\rm h}$34$^{\rm m}$26\fs6 & +60\degr 11\arcmin 13\farcs9& 13 & 22 & 17 & 9 & 0 & 0 & 48 & -0.13 & -0.31 & -1 & n/a \\
{\bf 5} &{\bf 20$^{\rm h}$34$^{\rm m}$34\fs4} & {\bf +60\degr 10\arcmin 31\farcs2}& {\bf 16} & {\bf 18} & {\bf 325} & {\bf 262} & {\bf 45} & {\bf 2} & {\bf 652} & {\bf 0.9} & {\bf -0.11} & {\bf -0.71} & {\bf -0.91} \\
{\bf 6} &{\bf 20$^{\rm h}$34$^{\rm m}$36\fs5} & {\bf +60\degr 09\arcmin 30\farcs5}& {\bf 21} & {\bf 160} & {\bf 723} & {\bf 520} & {\bf 79} & {\bf 42} & {\bf 1524} & {\bf 0.64} & {\bf -0.16} & {\bf -0.74} & {\bf -0.31} \\
7 &20$^{\rm h}$34$^{\rm m}$37\fs8 & +60\degr 04\arcmin 33\farcs1& 20 & 13 & 78 & 163 & 108 & 65 & 427 & 0.71 & 0.35 & -0.2 & -0.25 \\
8 &20$^{\rm h}$34$^{\rm m}$40\fs1 & +60\degr 08\arcmin 23\farcs3& 14 & 0 & 111 & 102 & 65 & 28 & 306 & 1 & -0.04 & -0.22 & -0.4 \\
9 &20$^{\rm h}$34$^{\rm m}$44\fs8 & +60\degr 07\arcmin 18\farcs0& 15 & 31 & 192 & 94 & 19 & 19 & 355 & 0.72 & -0.34 & -0.66 & 0 \\
10&20$^{\rm h}$34$^{\rm m}$46\fs6 & +60\degr 13\arcmin 25\farcs1& 15 & 71 & 52 & 62 & 69 & 69 & 323 & -0.15 & 0.09 & 0.05 & 0 \\
{\bf 11}&{\bf 20$^{\rm h}$34$^{\rm m}$48\fs0} & {\bf +60\degr 09\arcmin 32\farcs2}& {\bf 16} & {\bf 97} & {\bf 583} & {\bf 690} & {\bf 252} & {\bf 80} & {\bf 1702} & {\bf 0.71} & {\bf 0.08} & {\bf -0.46} & {\bf -0.52} \\
{\bf 12}&{\bf 20$^{\rm h}$34$^{\rm m}$48\fs8} & {\bf +60\degr 08\arcmin 10\farcs1}& {\bf 17} & {\bf 42} & {\bf 312} & {\bf 441} & {\bf 203} & {\bf 79} & {\bf 1077} & {\bf 0.76} & {\bf 0.17} & {\bf -0.37} & {\bf -0.44} \\
13&20$^{\rm h}$34$^{\rm m}$49\fs0 & +60\degr 12\arcmin 17\farcs0& 18 & 6 & 96 & 190 & 101 & 47 & 440 & 0.89 & 0.33 & -0.31 & -0.36 \\
{\bf 14}&{\bf 20$^{\rm h}$34$^{\rm m}$49\fs1} & {\bf +60\degr 05\arcmin 53\farcs8}& {\bf 19} & {\bf 106} & {\bf 599} & {\bf 194} & {\bf 21} & {\bf 9} & {\bf 929} & {\bf 0.7} & {\bf -0.51} & {\bf -0.8} & {\bf -0.39} \\
15&20$^{\rm h}$34$^{\rm m}$49\fs4 & +60\degr 13\arcmin 41\farcs9& 11 & 0 & 21 & 24 & 30 & 22 & 97 & 1 & 0.07 & 0.11 & -0.15 \\
{\bf 16}&{\bf 20$^{\rm h}$34$^{\rm m}$50\fs4} & {\bf +60\degr 10\arcmin 17\farcs0}& {\bf 18} & {\bf 79} & {\bf 608} & {\bf 470} & {\bf 177} & {\bf 70} & {\bf 1404} & {\bf 0.77} & {\bf -0.13} & {\bf -0.45} & {\bf -0.43} \\
{\bf 17}&{\bf 20$^{\rm h}$34$^{\rm m}$51\fs4} & {\bf +60\degr 10\arcmin 43\farcs7}& {\bf 17} & {\bf 123} & {\bf 871} & {\bf 1468} & {\bf 827} & {\bf 201} & {\bf 3490} & {\bf 0.75} & {\bf 0.26} & {\bf -0.28} & {\bf -0.61} \\
18&20$^{\rm h}$34$^{\rm m}$51\fs8 & +60\degr 12\arcmin 48\farcs6& 11 & 2 & 31 & 45 & 7 & 7 & 92 & 0.88 & 0.18 & -0.73 & 0 \\
{\bf 19}&{\bf 20$^{\rm h}$34$^{\rm m}$52\fs4} & {\bf +60\degr 09\arcmin 10\farcs6}& {\bf 23} & {\bf 269} & {\bf 2545} & {\bf 4542} & {\bf 2061} & {\bf 499} & {\bf 9916} & {\bf 0.81} & {\bf 0.28} & {\bf -0.38} & {\bf -0.61} \\
{\bf 20}&{\bf 20$^{\rm h}$34$^{\rm m}$53\fs5} & {\bf +60\degr 10\arcmin 54\farcs8}& {\bf 16} & {\bf 66} & {\bf 411} & {\bf 465} & {\bf 233} & {\bf 59} & {\bf 1234} & {\bf 0.72} & {\bf 0.06} & {\bf -0.33} & {\bf -0.6} \\
{\bf 21}&{\bf 20$^{\rm h}$34$^{\rm m}$56\fs7} & {\bf +60\degr 08\arcmin 32\farcs6}& {\bf 20} & {\bf 84} & {\bf 651} & {\bf 809} & {\bf 529} & {\bf 279} & {\bf 2352} & {\bf 0.77} & {\bf 0.11} & {\bf -0.21} & {\bf -0.31} \\
22&20$^{\rm h}$34$^{\rm m}$57\fs1 & +60\degr 05\arcmin 32\farcs6& 17 & 6 & 68 & 82 & 18 & 38 & 212 & 0.84 & 0.09 & -0.64 & 0.36 \\
{\bf 23}&{\bf 20$^{\rm h}$34$^{\rm m}$57\fs8} & {\bf +60\degr 09\arcmin 48\farcs0}& {\bf 14} & {\bf 33} & {\bf 446} & {\bf 1004} & {\bf 475} & {\bf 58} & {\bf 2016} & {\bf 0.86} & {\bf 0.38} & {\bf -0.36} & {\bf -0.78} \\
{\bf 24}&{\bf 20$^{\rm h}$35$^{\rm m}$00\fs2} & {\bf +60\degr 09\arcmin 08\farcs0}& {\bf 24} & {\bf 812} & {\bf 4413} & {\bf 3982} & {\bf 688} & {\bf 99} & {\bf 9994} & {\bf 0.69} & {\bf -0.05} & {\bf -0.71} & {\bf -0.75} \\
25&20$^{\rm h}$35$^{\rm m}$00\fs7 & +60\degr 12\arcmin 50\farcs8& 15 & 0 & 34 & 53 & 31 & 7 & 125 & 1 & 0.22 & -0.26 & -0.63 \\
{\bf 26}&{\bf 20$^{\rm h}$35$^{\rm m}$00\fs8} & {\bf +60\degr 11\arcmin 29\farcs9}& {\bf 38} & {\bf 3684} & {\bf 16674} & {\bf 17157} & {\bf 5547} & {\bf 1588} & {\bf 44650} & {\bf 0.64} & {\bf 0.01} & {\bf -0.51} & {\bf -0.55} \\
{\bf 27}&{\bf 20$^{\rm h}$35$^{\rm m}$01\fs2} & {\bf +60\degr 10\arcmin 09\farcs4}& {\bf 18} & {\bf 87} & {\bf 614} & {\bf 1017} & {\bf 638} & {\bf 263} & {\bf 2619} & {\bf 0.75} & {\bf 0.25} & {\bf -0.23} & {\bf -0.42} \\
28&20$^{\rm h}$35$^{\rm m}$05\fs4 & +60\degr 06\arcmin 21\farcs4& 16 & 7 & 50 & 160 & 115 & 63 & 395 & 0.75 & 0.52 & -0.16 & -0.29 \\
29&20$^{\rm h}$35$^{\rm m}$11\fs5 & +60\degr 08\arcmin 57\farcs8& 15 & 33 & 194 & 133 & 13 & 2 & 375 & 0.71 & -0.19 & -0.82 & -0.73 \\
30&20$^{\rm h}$35$^{\rm m}$12\fs7 & +60\degr 07\arcmin 31\farcs4& 16 & 0 & 55 & 83 & 53 & 9 & 200 & 1 & 0.2 & -0.22 & -0.71 \\
31&20$^{\rm h}$35$^{\rm m}$14\fs0 & +60\degr 05\arcmin 21\farcs0& 16 & 15 & 43 & 74 & 55 & 34 & 221 & 0.49 & 0.26 & -0.15 & -0.24 \\
{\bf 32}&{\bf 20$^{\rm h}$35$^{\rm m}$18\fs7} & {\bf +60\degr 10\arcmin 55\farcs5}& {\bf 18} & {\bf 32} & {\bf 183} & {\bf 767} & {\bf 445} & {\bf 114} & {\bf 1541} & {\bf 0.7} & {\bf 0.61} & {\bf -0.27} & {\bf -0.59} \\
33&20$^{\rm h}$35$^{\rm m}$25\fs5 & +60\degr 09\arcmin 54\farcs3& 18 & 9 & 221 & 79 & 7 & 2 & 318 & 0.92 & -0.47 & -0.84 & -0.6 \\
34&20$^{\rm h}$35$^{\rm m}$28\fs1 & +60\degr 09\arcmin 20\farcs5& 14 & 3 & 221 & 79 & 7 & 0 & 310 & 0.97 & -0.47 & -0.84 & -1 \\
35&20$^{\rm h}$35$^{\rm m}$32\fs4 & +60\degr 10\arcmin 17\farcs9& 19 & 18 & 68 & 50 & 48 & 11 & 195 & 0.57 & -0.15 & -0.02 & -0.63 \\
\hline
\end{tabular}
\tablefoot{(1)Number as in Fig.~\ref{6946points}, (2),(3) equatorial coordinates, (4) extraction radius in arcsec, (5)-(9) net
counts in relevant energy bands (see Sect.~\ref{pointdist}),
(10) net counts in the total energy band, (11)-(14) hardness ratios (see Sect.~\ref{pointdist}). Sources for which spectra were
fitted with models (see Sect.~\ref{spectra}) are listed in boldface.}
\end{sidewaystable*}

The left panel of Fig.~\ref{6946points} suggests that some of the sources might not originate from within the galaxy and are instead more distant background objects. 
Moreover, the centre of the most extended
region of source 26 does not seem to coincide with the visible H$\alpha$ clump, but more precise astrometry of the Chandra observation
associated this source with the galaxy \citep[e.g.][]{kajava09}. 
This source is the brightest point source -- the ultra-luminous complex MF16 that is assumed 
to be a supernova remnant \citep[e.g.][]{matonick}.
It is not a pure point source, as it is considerably larger than the point spread function
of the XMM-Newton EPIC cameras (of $\simeq$ 12\arcsec).

Another interesting source is source 17, overlapping in Fig.~\ref{6946points} with sources 16 and 20.
This source was not visible in previous observations (where
sources 16 and 20 were detected), but only in the most recent sensitive observations.

\subsection{Spectral analysis of the X-ray emission}
\label{spectra}

For spectra of the hot gas we used a model that attributed one or two thermal plasmas 
and/or a contribution from unresolved point-like sources.
Thermal plasma is represented in this work by a {\it mekal} model, which is a model of an emission spectrum from hot diffuse gas
based on the model calculations of Mewe and Kaastra \citep{mewe,kaastra}. In all models we fixed the
metallicity to solar. This was determined by a test spectral fitting of the spectra, which showed that in a wide range of metallicities (0.1 - 1.5 solar), 
no significant change in gas temperatures or power-law photon indices could be observed. Slight deviations of these parameters were still within 
the errors provided by the models that used solar metallicity. This suggests that all differences that are found in gas temperatures when fitting 
spectra do not result from a possible combined effect of abundance gradients and enforcing solar metallicities in the models. It also shows that an additional power-law 
component in the models of diffuse emission does account for emission from unresolved point-like sources and is not introduced to mimic variable gas abundance because excluding the power-law component from the model and again varying metallicities resulted in unphysical values of the parameters. Consequently, the quality of the 
model fits remained very low (with reduced $\chi^2 > 2$).

A contribution from unresolved point-like sources is fitted with a simple power law.
For some models an additional component to account for the internal absorption needed to be used. For all models we also used a
fixed foreground (galactic) absorption.
Tables 4 and 8 present an overview of the various spectral models, including the obtained parameters. The errors provided for the model parameters are always 1-$\sigma$ errors.
The derived X-ray fluxes and luminosities are shown in Tables~\ref{6946xfs} and~\ref{6946xfr}.

The same model components were used for the spectral analysis of the brightest point sources (see Sect.~\ref{pointdist}). Since we did not aim to perform a variability analysis of point sources,  we used all data that were suitable for spectral analysis 
(for the selection and merging of separate data sets, see Sect.~\ref{obsred}),
similarly as for regions of diffuse emission.

The regions of the diffuse X-ray emission from which the spectra were extracted are presented in Fig.~\ref{6946xregs}.
Extraction areas for the brightest point-sources are the same as found by the source-detection analysis (cf. Fig.~\ref{6946points}).
All plots of the modelled spectra together with the fitted models are presented in Figs.~\ref{6946mod1} and \ref{6946mod2}.

\subsubsection{Brightest sources}
\label{sourcespec}

For almost all sources (except for source 27) for which spectral fitting was performed, we also used the thermal plasma component (see Table~\ref{6946xtabs}) 
to account for diffuse emission projected onto each of the sources. The amount of this emission was often significant 
as a result of the sensitive observations of a gas-rich galaxy with a relatively large point spread function of the EPIC-pn camera that did not allow clearly separating the studied sources.
Many of the spectra also required an additional model component to account for the high internal absorption.

For one source (source 5) a model with a power-law component did not give good results. The spectrum was fitted best when using two thermal
components. It is possible that it is simply a hot gas clump that was classified by the source detection routine to be a point
source.
Alternatively, the lowest count number of all fitted spectra might also be the reason for difficulties in finding a proper fit. The
hot component fitted in this model reaches values of almost 2\,keV, which is fairly unexpected for the hot gas in the galactic disk.

For source 17, which was undetected in earlier observations, a complex model had to be fitted. Its power-law component is poorly constrained, however.
The two sources near this position that were detected previously,
sources 16 and 20, were not detected in the recent sensitive observation because the brightness of the diffuse emission was
higher throughout, and also because the bright source 17 has wings. For consistency of the results, however, we used all observations to extract spectra for all three sources. Nevertheless, the complexity
of the model for source 17 might be caused by admixtures from the two adjacent sources.

The spectrum of the core region, source 19, where the densest and the hottest gas might be expected, could be fitted equally well with either one or two thermal components.
We present both model fits. The model with two components yielded 0.32$^{+0.02}_{-0.03}$\,keV and 0.70$^{+0.07}_{-0.06}$\,keV, while a single temperature component
showed a temperature of 0.62$\pm$0.04\,keV. These values and the derived fluxes for this region (see Table~\ref{6946xfs}) suggest
that most of the emission originates from the hotter component. The main difference between both model fits, however, is a significant variation
of the internal absorption (see Table~\ref{6946xtabs}).

By fitting a model (consisting of one power-law
and two thermal components) to the spectrum of MF16 (source 26), we obtained a temperature of 1.04$\pm$0.03\,keV for
the \textup{{\em \textup{hot}}} component.
Still, most of the flux from MF16 originates from the power-law component that is associated with the central source. To further investigate the spectrum of this source, we followed
\citet{kajava09} and added a multicolour disk component to our
model. We obtained a similar value for the temperature of the inner disk.

\begin{table*}[ht]
\caption{\label{6946xtabs}Model-fit parameters for the point sources studied in NGC\,6946.}
\centering
\begin{tabular}{llccccrr}
\hline\hline
ID                      & Model                           &       Internal                & kT$_1$                        & kT$_2$                        &       Photon                  &$\chi_{\rm red}^2$& Net   \\
                        & type                            &       nH\tablefootmark{a}     & [keV]                         & [keV]                         &       Index                   &                 & counts\\
\hline
\vspace{5pt}
3                       & wabs(mekal+wabs*power law)       & 4.19$^{+2.07}_{-1.76}$        & 0.23$^{+0.04}_{-0.05}$        & --                            & 2.62$\pm$0.23                 & 0.94             & 1368  \\
\vspace{5pt}
5                       & wabs(mekal+mekal)               & --                            & 0.54$\pm$0.09                 & 1.93$^{+0.65}_{-0.32}$        & --                            & 0.80      &  652  \\
\vspace{5pt}
6                       & wabs(mekal+wabs*power law)       & 1.86$^{+0.47}_{-0.44}$        & 0.60$^{+0.11}_{-0.35}$        & --                            & 3.91$^{+0.61}_{-0.41}$        & 1.04             & 1524  \\
\vspace{5pt}
11                      & wabs(mekal+wabs*power law)       & 1.78$^{+1.66}_{-0.54}$        & 0.38$^{+0.16}_{-0.08}$        & --                            & 2.48$^{+0.48}_{-0.21}$        & 0.99             & 1702  \\
\vspace{5pt}
12                      & wabs(mekal+power law)            & --                            & 0.58$^{+0.21}_{-0.22}$        & --                            & 1.78$\pm$0.12                 & 0.91      & 1077  \\
\vspace{5pt}
14                      & wabs(mekal+power law)            & --                            & 0.46$^{+0.06}_{-0.08}$        & --                            & 3.12$^{+0.56}_{-0.91}$        & 1.12      &  929  \\
\vspace{5pt}
16                      & wabs(mekal+power law)            & --                            & 0.38$^{+0.21}_{-0.06}$        & --                            & 2.03$^{+0.15}_{-0.13}$        & 1.15      & 1404  \\
\vspace{5pt}
17                      & wabs(mekal+mekal+wabs*power law) & 5.29$^{+2.05}_{-1.33}$        & 0.18$\pm$0.02                 & 0.66$\pm$0.06                 & 2.31$^{+0.83}_{-1.04}$        & 0.91             & 3490  \\
\vspace{5pt}
19                      & wabs(mekal+wabs(mekal+power law))& 6.87$^{+1.02}_{-1.21}$        & 0.32$^{+0.02}_{-0.03}$        & 0.70$^{+0.07}_{-0.06}$        & 2.55$^{+0.07}_{-0.03}$        & 1.02        & 9916  \\
\vspace{5pt}
19\tablefootmark{b}     & wabs(mekal+wabs*power law)       & 2.89$^{+0.31}_{-0.33}$        & 0.62$\pm$0.04                 & --                            & 2.36$^{+0.03}_{-0.07}$        & 1.03             & 9916  \\
\vspace{5pt}
20                      & wabs(mekal+power law)            & --                            & 0.62$^{+0.08}_{-0.12}$        & --                            & 1.65$^{+0.12}_{-0.11}$        & 0.90      & 1234  \\
\vspace{5pt}
21                      & wabs(mekal+power law)            & --                            & 0.45$^{+0.08}_{-0.06}$        & --                            & 1.52$^{+0.08}_{-0.07}$        & 1.12      & 2352  \\
\vspace{5pt}
23                      & wabs(mekal+wabs*power law)       & 3.29$^{+0.81}_{-1.05}$        & 0.30$^{+0.21}_{-0.07}$        & --                            & 2.52$^{+0.22}_{-0.20}$        & 1.10             & 2016  \\
\vspace{5pt}
24                      & wabs(mekal+wabs*power law)       & 2.16$^{+0.17}_{-0.21}$        & 0.78$\pm$0.05                 & --                            & 3.80$\pm$0.15                 & 1.23             & 9994  \\
\vspace{5pt}
26                      & wabs(mekal+mekal+power law)      & --                            & 0.40$^{+0.09}_{-0.03}$        & 1.04$\pm$0.03                 & 2.26$\pm$0.03                 & 1.19      & 44650 \\
\vspace{5pt}
26\tablefootmark{c}     &wabs(mekal+mekal+power law+diskbb)& --                            & 0.66$\pm$0.06                 & 1.34$^{+0.32}_{-0.22}$        & 1.98$\pm$0.08                 & 1.11      & 44650 \\
\vspace{5pt}
27                      & wabs*power law                   & --                            & --                            & --                            & 1.68$\pm$0.05                 & 1.05      & 2619  \\
\vspace{5pt}
32                      & wabs(mekal+wabs*power law)       & 8.83$^{+2.15}_{-1.70}$        & 0.25$^{+0.06}_{-0.05}$        & --                            & 2.82$^{+0.26}_{-0.17}$        & 0.80             & 1541  \\
\hline
\end{tabular}
\tablefoot{
\tablefoottext{a}{Column density in [10$^{21}$ cm$^{-2}$].}
\tablefoottext{b}{Single temperature model for the thermal component.}
\tablefoottext{c}{Model with an additional multicolour disk model component of $T_{in} = 0.26^{+0.02}_{-0.01}$\,keV.}
ID\,19 is the core central source in NGC\,6946 and ID\,26 is MF16.
}
\end{table*}

\begin{figure*}[ht]
\begin{center}
\resizebox{0.45\hsize}{!}{\includegraphics[angle=-90]{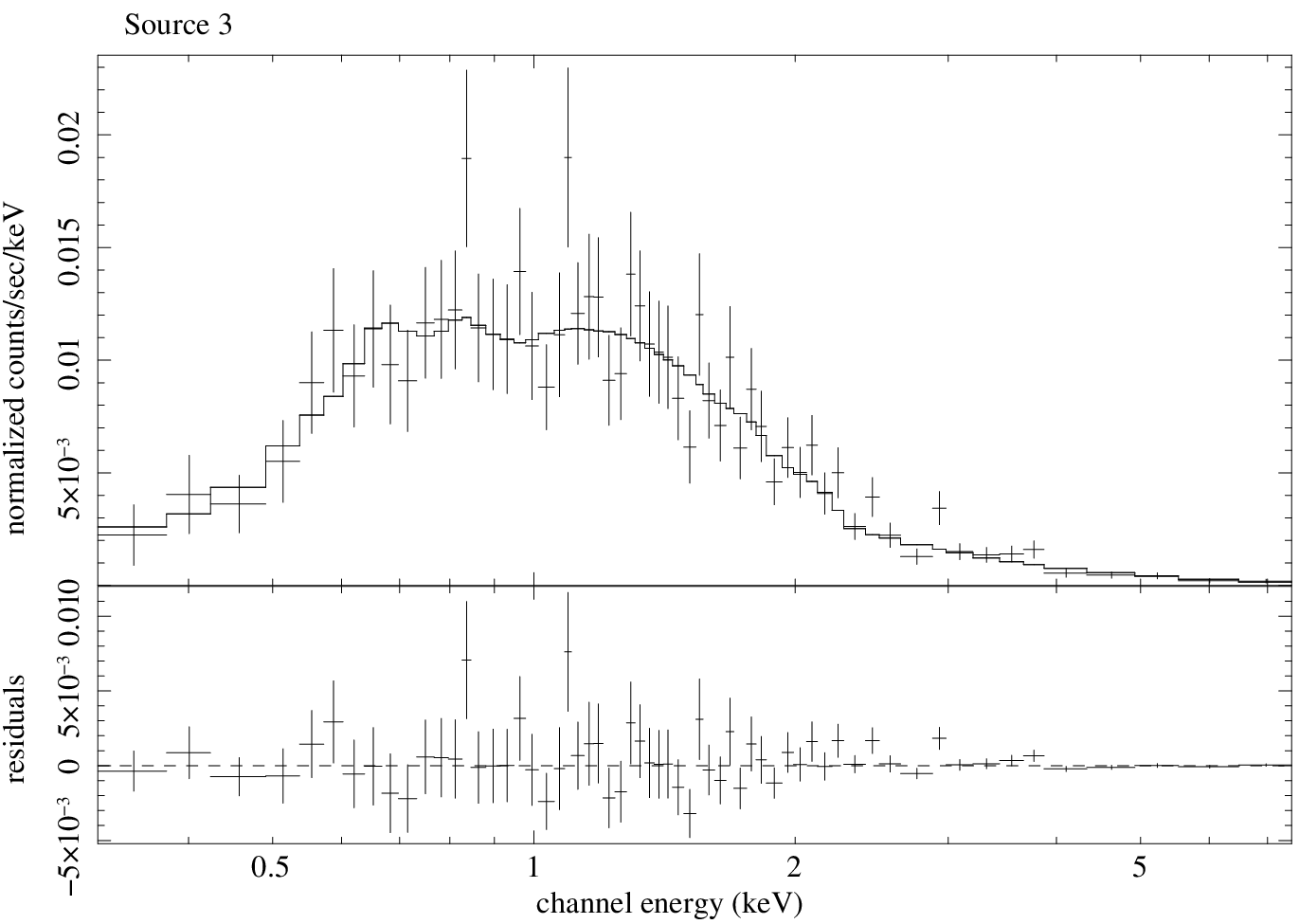}}
\resizebox{0.45\hsize}{!}{\includegraphics[angle=-90]{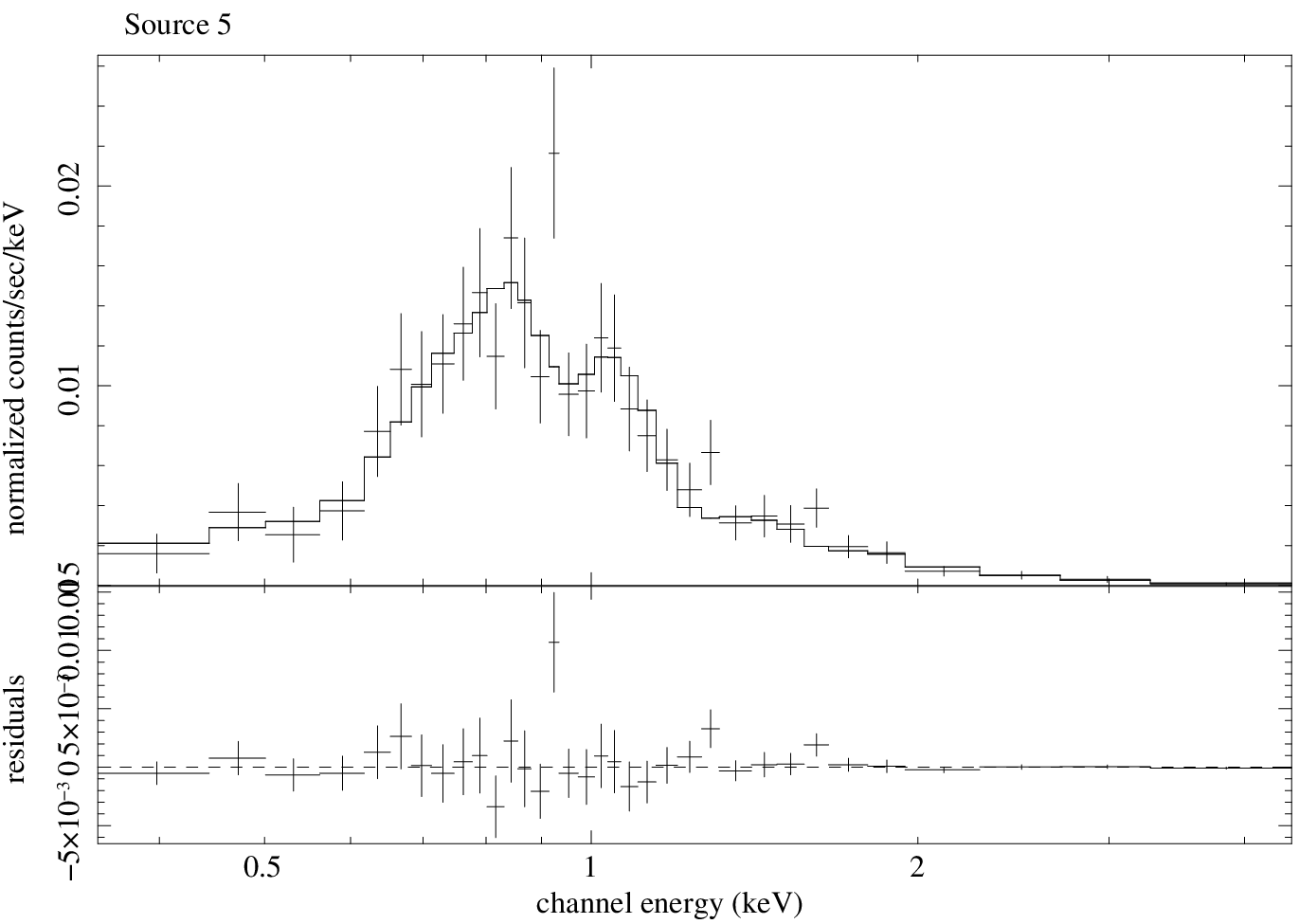}}
\resizebox{0.45\hsize}{!}{\includegraphics[angle=-90]{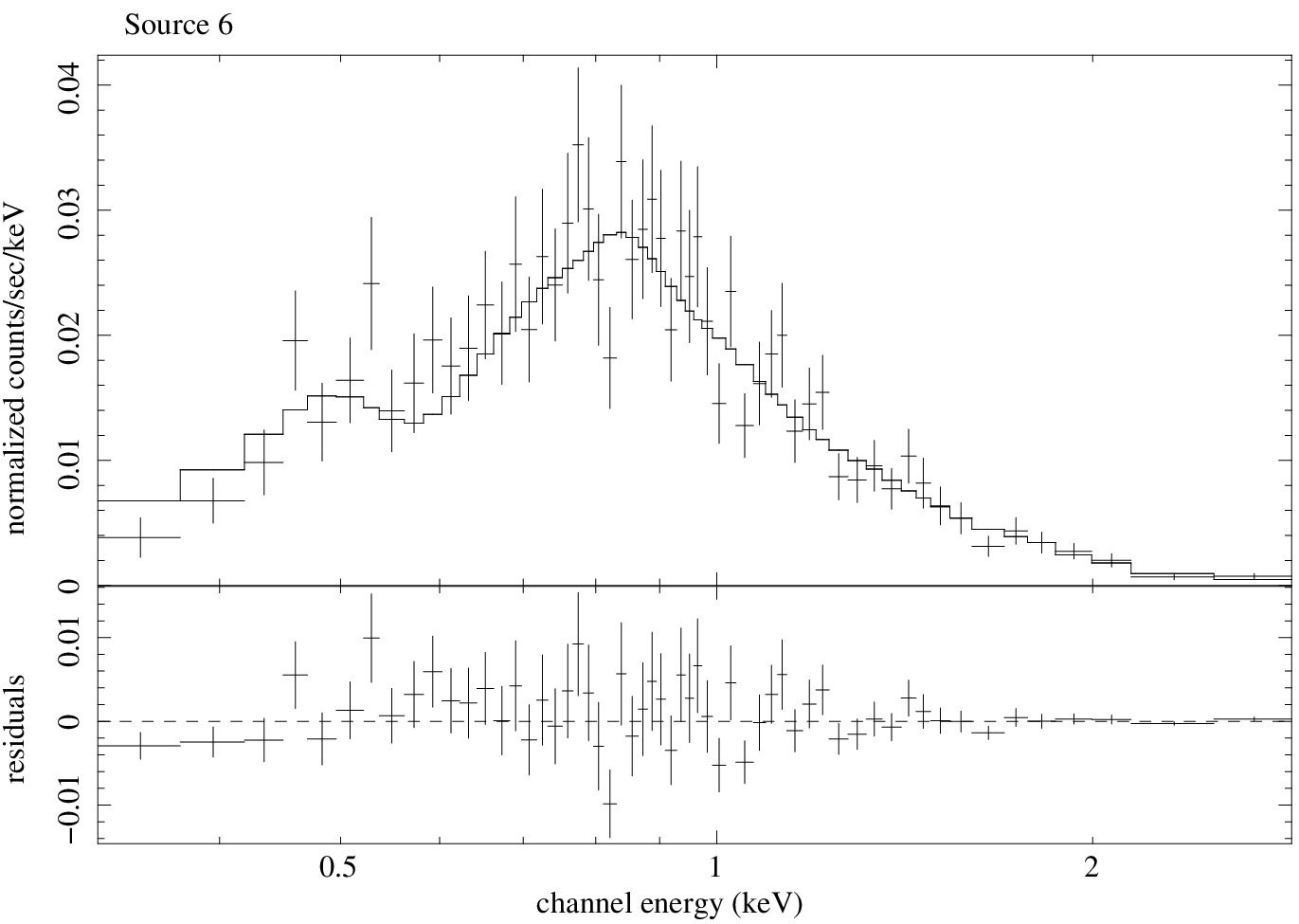}}
\resizebox{0.45\hsize}{!}{\includegraphics[angle=-90]{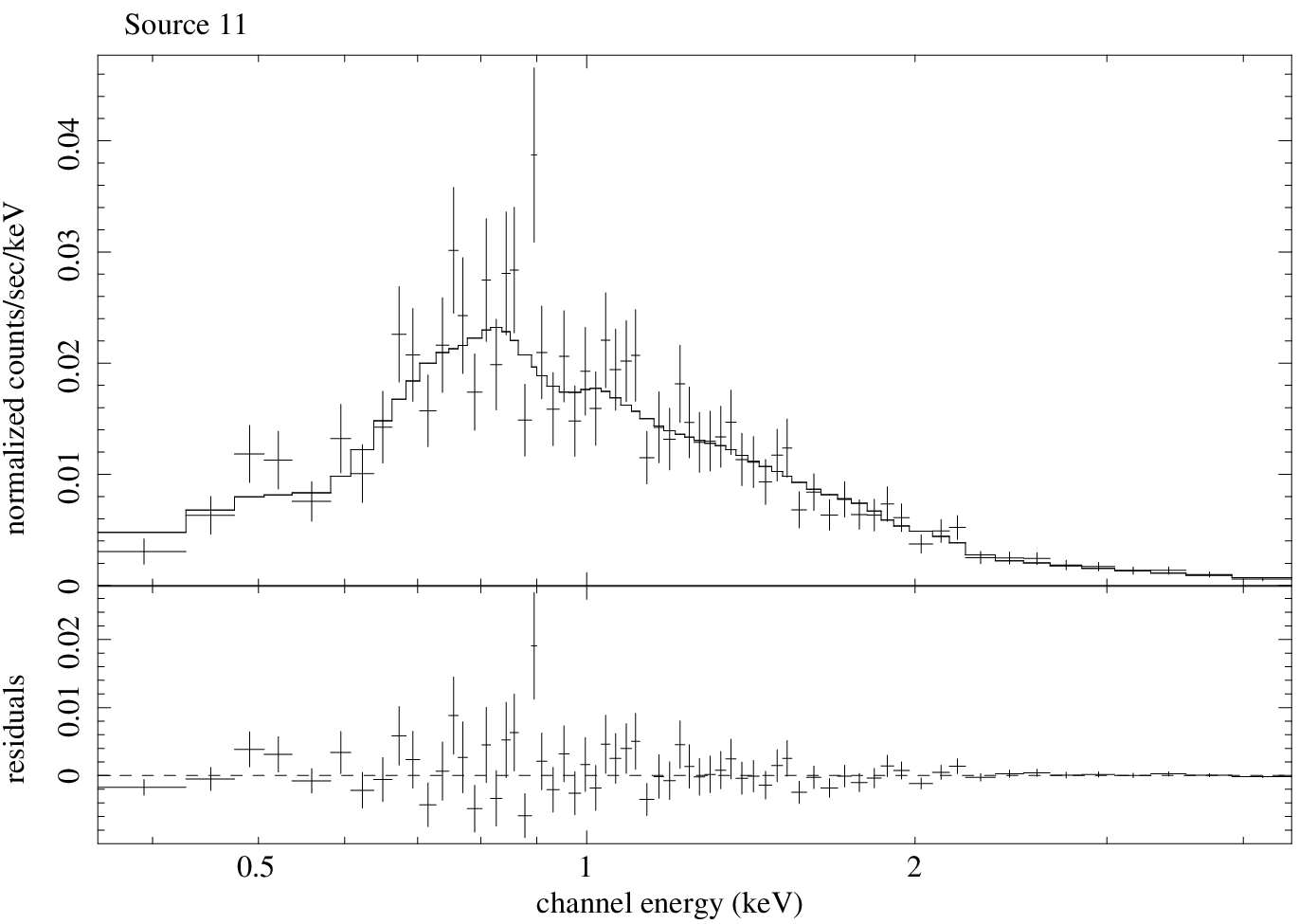}}
\resizebox{0.45\hsize}{!}{\includegraphics[angle=-90]{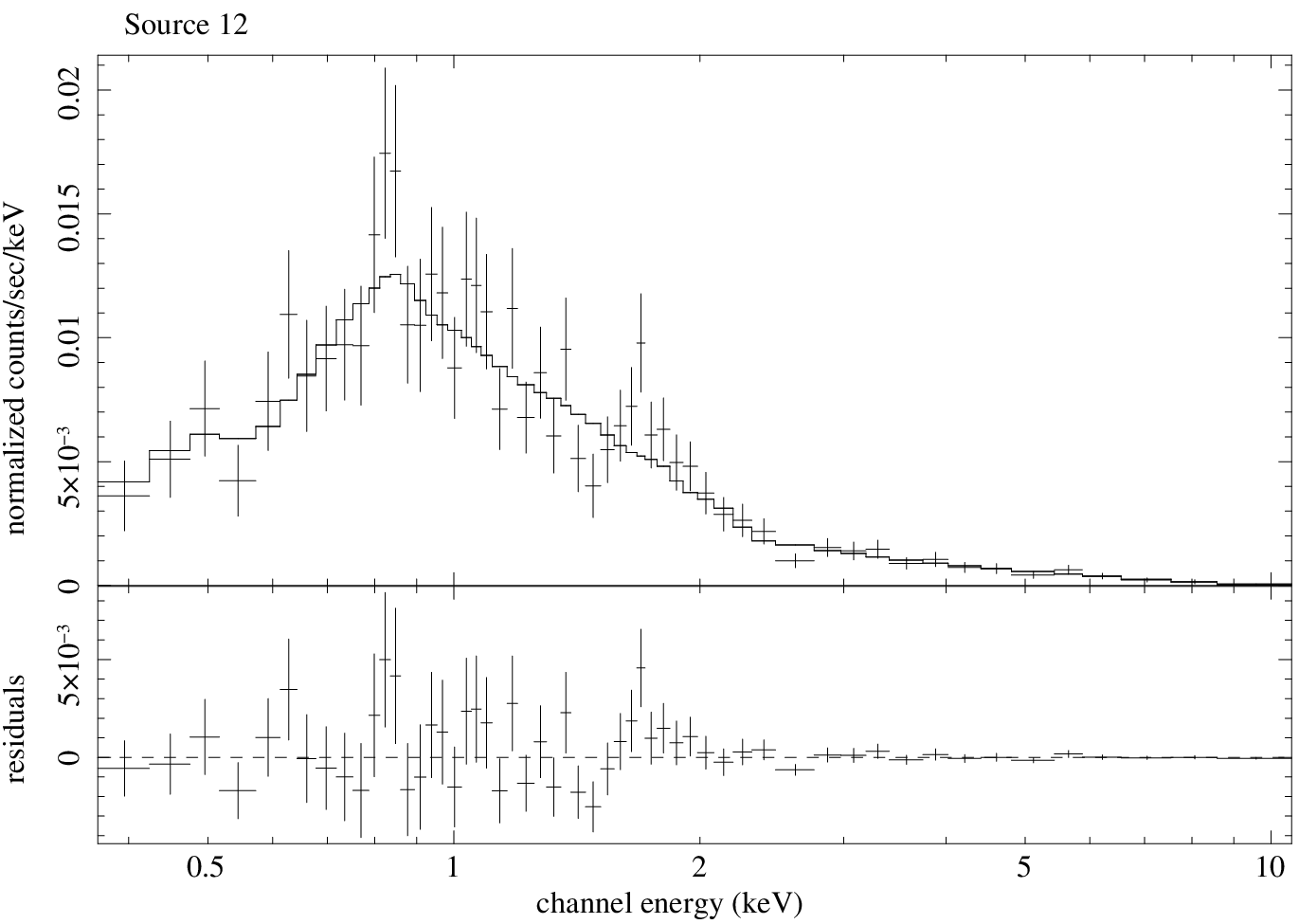}}
\resizebox{0.45\hsize}{!}{\includegraphics[angle=-90]{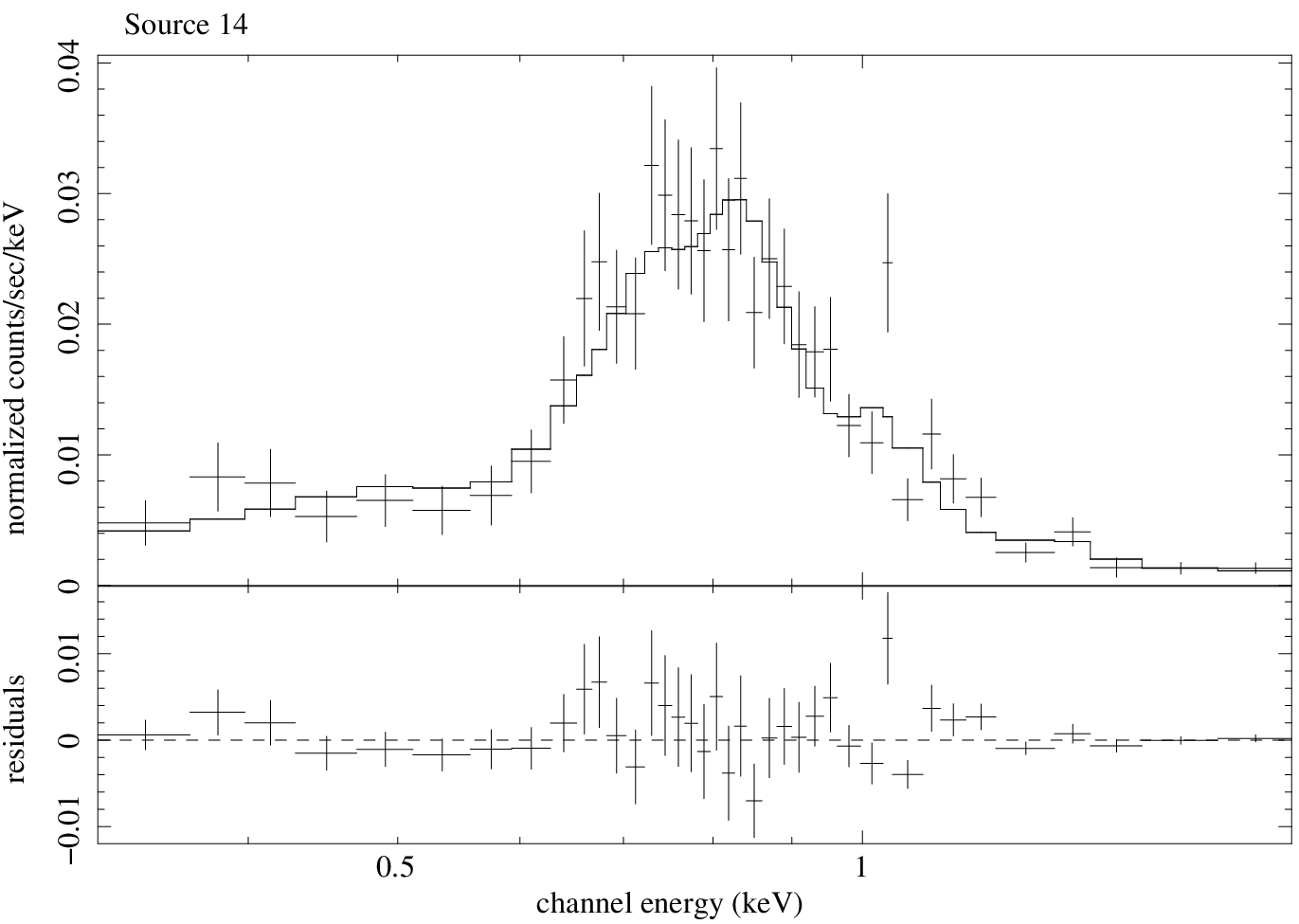}}
\resizebox{0.45\hsize}{!}{\includegraphics[angle=-90]{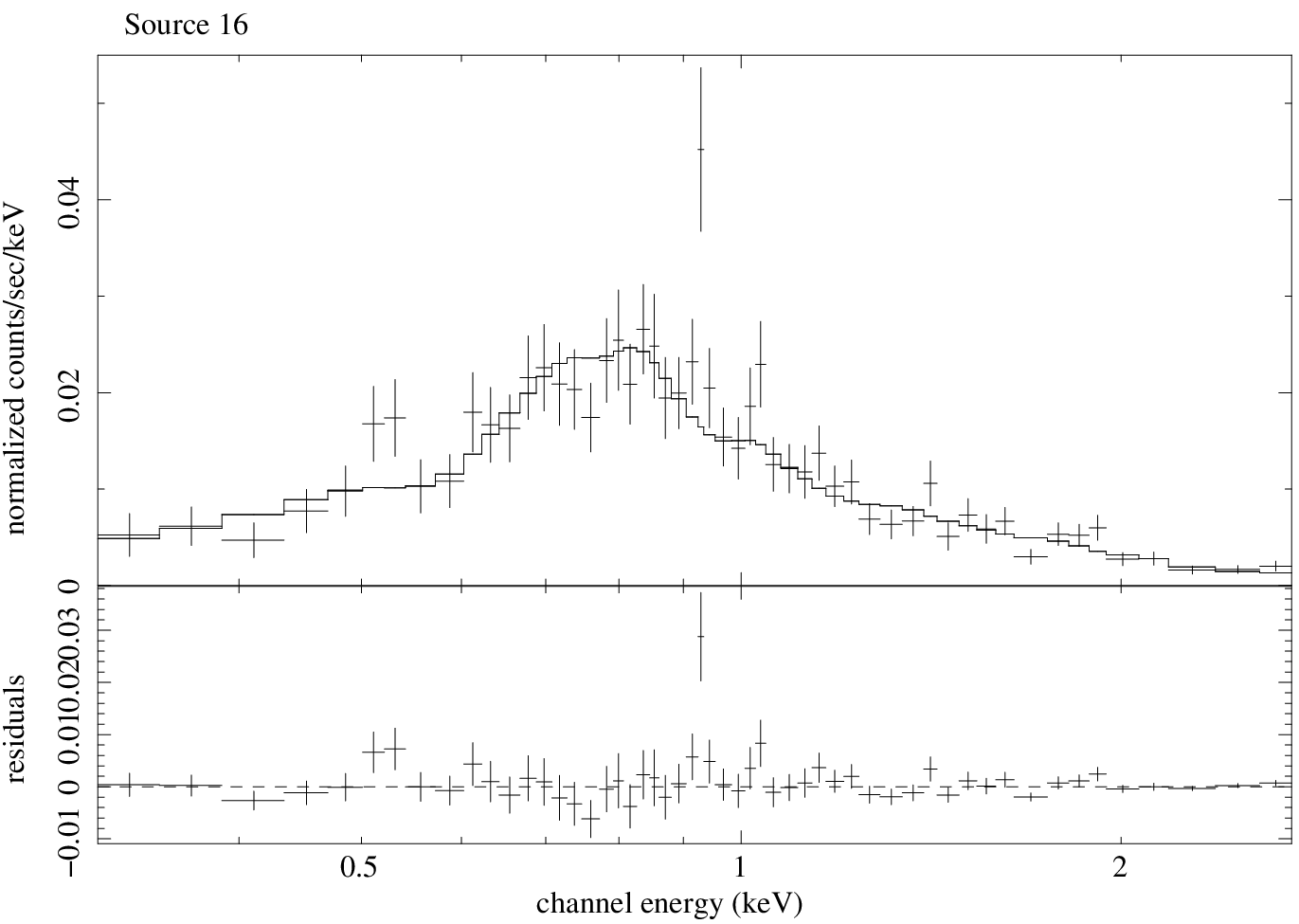}}
\resizebox{0.45\hsize}{!}{\includegraphics[angle=-90]{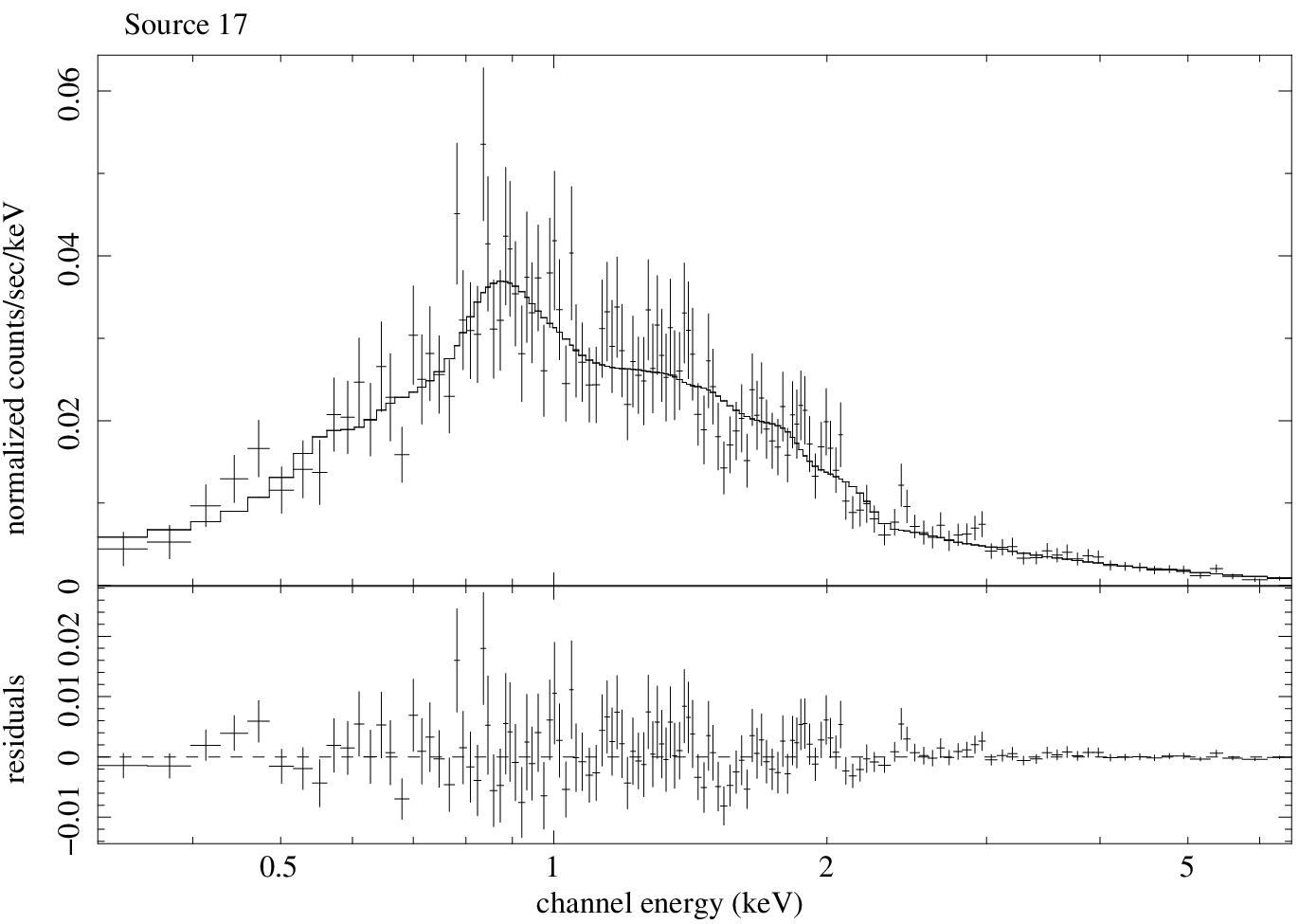}}
\end{center}
\caption{Model fits to the spectra of selected sources in NGC\,6946. See text and Tables~\ref{6946xtabs} and
\ref{6946xfs}.}
\label{6946mod3}
\end{figure*}

\begin{figure*}[ht]
\begin{center}
\resizebox{0.45\hsize}{!}{\includegraphics[angle=-90]{ngc6946_source_20.ps}}
\resizebox{0.45\hsize}{!}{\includegraphics[angle=-90]{ngc6946_source_21.ps}}
\resizebox{0.45\hsize}{!}{\includegraphics[angle=-90]{ngc6946_source_23.ps}}
\resizebox{0.45\hsize}{!}{\includegraphics[angle=-90]{ngc6946_source_24.ps}}
\resizebox{0.45\hsize}{!}{\includegraphics[angle=-90]{ngc6946_source_26.ps}}
\resizebox{0.45\hsize}{!}{\includegraphics[angle=-90]{ngc6946_source_26_dbb.ps}}
\resizebox{0.45\hsize}{!}{\includegraphics[angle=-90]{ngc6946_source_27.ps}}
\resizebox{0.45\hsize}{!}{\includegraphics[angle=-90]{ngc6946_source_32.ps}}
\end{center}
\caption{Model fits to the spectra of selected sources in NGC\,6946. See text and Tables~\ref{6946xtabs} and
\ref{6946xfs}.}
\label{6946mod4}
\end{figure*}

\begin{figure*}[ht]
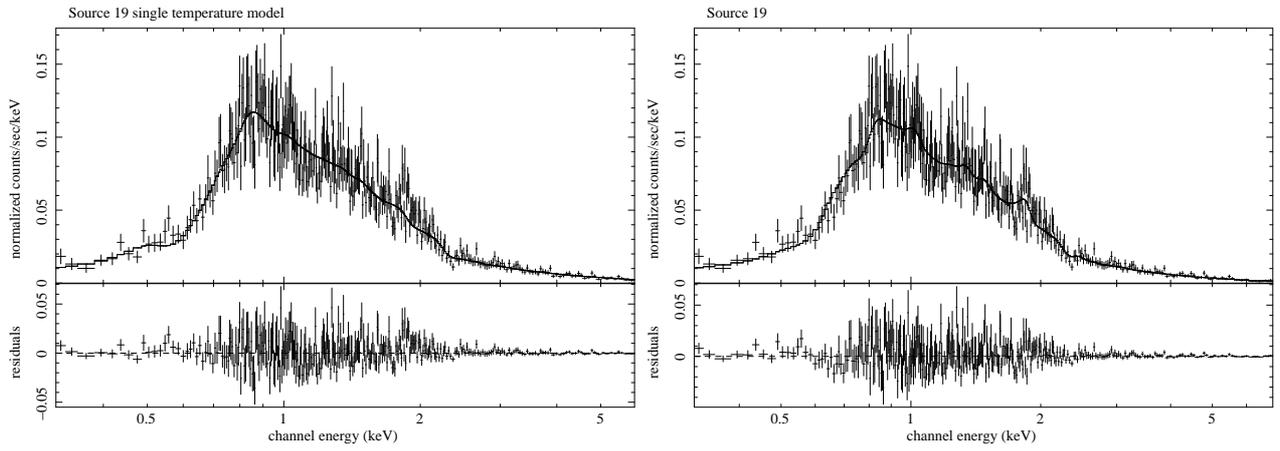

\begin{center}
\resizebox{0.45\hsize}{!}{\includegraphics[angle=-90]{ngc6946_source19_onetemp.ps}}
\resizebox{0.45\hsize}{!}{\includegraphics[angle=-90]{ngc6946_source_19.ps}}
\end{center}
\caption{Model fits to the spectra of the nuclear region of NGC\,6946. See text and Tables~\ref{6946xtabs} and \ref{6946xfs}.}
\label{6946nuclear}
\end{figure*}

\begin{figure*}[ht]
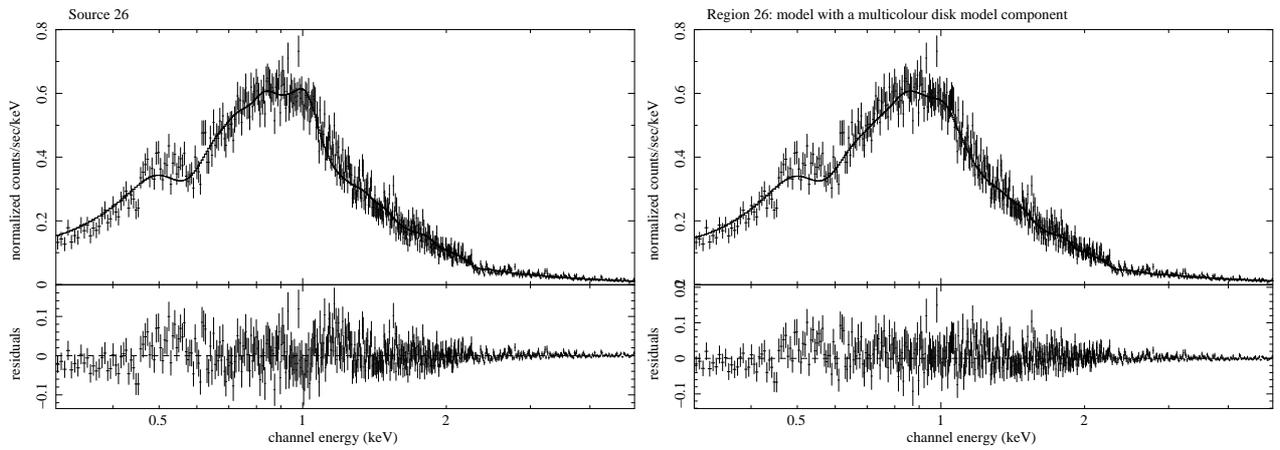

\begin{center}
\resizebox{0.45\hsize}{!}{\includegraphics[angle=-90]{ngc6946_source_26.ps}}
\resizebox{0.45\hsize}{!}{\includegraphics[angle=-90]{ngc6946_source_26_dbb.ps}}
\end{center}
\caption{Model fits to the spectra of MF16. See text and Tables~\ref{6946xtabs} and \ref{6946xfs}.}
\label{6946mf16}
\end{figure*}

\begin{table*}[ht]
        \caption{\label{6946xfs}Total (0.3 - 12 keV) unabsorbed fluxes in 10$^{-14}$erg\,cm$^{-2}$s$^{-1}$ for modelled point sources in NGC\,6946 along with luminosities
                                of their power-law components in 10$^{39}$erg\,s$^{-1}$.}
\centering
\begin{tabular}{cccccc}
\hline\hline
\vspace{5pt} ID         & mekal cold                    & mekal hot                     & power law                      & total                   &   luminosity   \\
\hline
3                       & 1.5$^{+0.6}_{-0.8}$           & --                            & 18.8$^{+6.3}_{-4.3}$          & 20.3$^{+6.8}_{-5.1}$    &  1.10$^{+0.37}_{-0.25}$    \\
\vspace{5pt}
5                       & 1.0$^{+0.1}_{-0.2}$           & 1.1$^{+0.3}_{-0.2}$           & --                            & 2.1$\pm$0.4             &  --    \\
\vspace{5pt}
6                       & 2.0$^{+0.7}_{-1.1}$           & --                            & 41.1$^{+31.3}_{-14.0}$        & 43.1$^{+32.0}_{-15.2}$  &  2.41$^{+1.84}_{-0.82}$    \\
\vspace{5pt}
11                      & 0.8$^{+0.9}_{-0.4}$           & --                            & 9.7$^{+4.9}_{-2.5}$           & 10.5$^{+5.8}_{-2.9}$    &  0.57$^{+0.29}_{-0.15}$    \\
\vspace{5pt}
12                      & 0.3$\pm$0.2                   & --                            & 5.3$^{+1.3}_{-1.2}$           & 5.6$^{+1.5}_{-1.4}$     &  0.31$^{+0.08}_{-0.07}$    \\
\vspace{5pt}
14                      & 2.0$^{+0.8}_{-0.4}$           & --                            & 1.3$^{+1.0}_{-0.6}$           & 3.3$^{+1.1}_{-0.7}$     &  0.08$^{+0.06}_{-0.04}$    \\
\vspace{5pt}
16                      & 1.2$^{+0.5}_{-0.4}$           & --                            & 4.3$^{+1.0}_{-0.8}$           & 5.5$^{+1.6}_{-1.3}$     &  0.25$^{+0.06}_{-0.05}$    \\
\vspace{5pt}
17                      & 1.6$^{+0.6}_{-0.5}$           & 1.6$\pm$0.6                   & 25.7$^{+5.2}_{-5.8}$          & 28.9$^{+6.3}_{-6.9}$    &  1.51$^{+0.30}_{-0.34}$    \\
\vspace{5pt}
19                      & 4.9$^{+0.6}_{-0.8}$           & 26.0$^{+15.6}_{-11.2}$        & 69.6$^{+6.5}_{-2.9}$          & 100.5$^{+22.8}_{-14.9}$ &  4.08$^{+0.38}_{-0.17}$    \\
\vspace{5pt}
19\tablefootmark{a}     & 3.6$\pm$0.6                   & --                            & 58.6$^{+5.7}_{-2.4}$          & 62.2$^{+6.3}_{-3.0}$    &  3.44$^{+0.33}_{-0.14}$    \\
\vspace{5pt}
20                      & 0.8$\pm$0.2                   & --                            & 5.6$^{+1.6}_{-1.4}$           & 6.4$^{+1.7}_{-1.5}$     &  0.33$^{+0.09}_{-0.08}$    \\
\vspace{5pt}
21                      & 0.9$\pm$0.3                   & --                            & 11.4$^{+2.1}_{-1.9}$          & 12.4$^{+2.3}_{-2.2}$    &  0.67$^{+0.12}_{-0.11}$    \\
\vspace{5pt}
23                      & 0.6$^{+0.7}_{-0.4}$           & --                            & 17.7$^{+5.3}_{-3.3}$          & 18.3$^{+5.9}_{-3.8}$    &  1.04$^{+0.31}_{-0.19}$    \\
\vspace{5pt}
24                      & 4.5$\pm$0.9                   & --                            & 108.0$^{+16.7}_{-14.9}$       & 112.6$^{+17.6}_{-15.9}$ &  6.33$^{+0.98}_{-0.87}$    \\
\vspace{5pt}
26                      & 10.5$^{+2.5}_{-2.0}$          & 14.5$^{+1.7}_{-1.6}$          & 128.3$^{+4.2}_{-4.4}$         & 153.3$^{+8.4}_{-8.0}$   &  7.52$^{+0.25}_{-0.26}$    \\
\vspace{5pt}
26\tablefootmark{b}     & 10.2$\pm$2.4                  & 9.4$^{+2.3}_{-1.4}$           & 104.1$^{+14.4}_{-16.9}$       & 156.2$^{+32.1}_{-21.2}$ &  8.03$^{+2.05}_{-1.76}$    \\
\vspace{5pt}
27                      & --                            & --                            & 14.8$^{+1.4}_{-1.3}$          & 14.8$^{+1.4}_{-1.3}$    &  0.87$^\pm$0.08    \\
\vspace{5pt}
32                      & 0.7$\pm$0.3                   & --                            & 36.3$^{+9.3}_{-6.0}$          & 37.0$^{+9.6}_{-6.2}$    &  2.13$^{+0.55}_{-0.35}$    \\
\hline
\end{tabular}
\tablefoot{
\tablefoottext{a}{Single temperature model for the thermal component.}
\tablefoottext{b}{Model with an additional multicolour disk model component with a flux of 32.9$^{+21.4}_{-12.8}$$\times$10$^{-14}$erg\,cm$^{-2}$s$^{-1}$.}
ID\,19 is the core central source in NGC\,6946 and ID\,26 is MF16.
}
\end{table*}

\subsubsection{Regions of diffuse emission}
\label{regspec}

\begin{figure*}[ht]
\begin{center}
\resizebox{0.46\hsize}{!}{\includegraphics[clip]{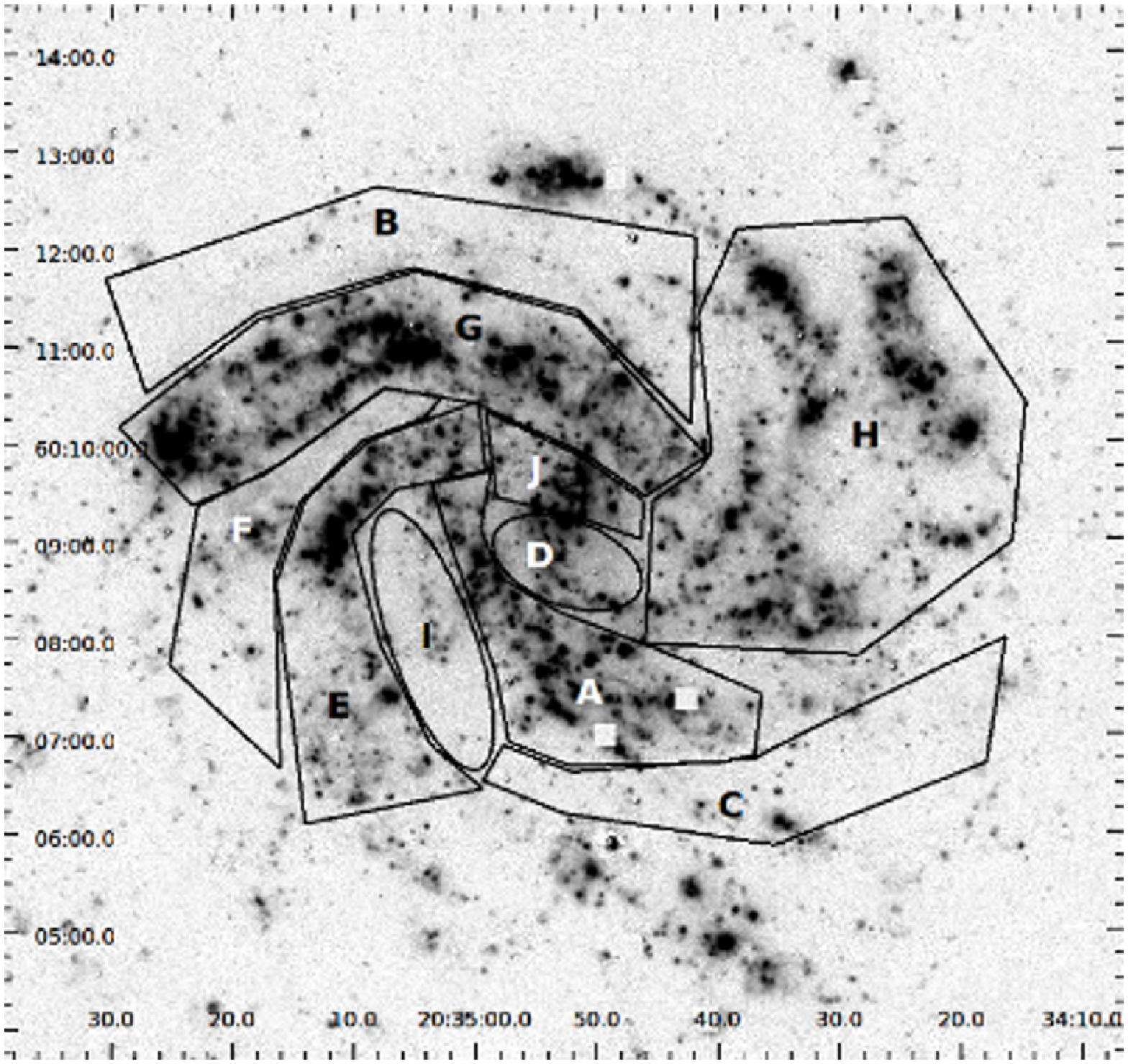}}
\resizebox{0.45\hsize}{!}{\includegraphics{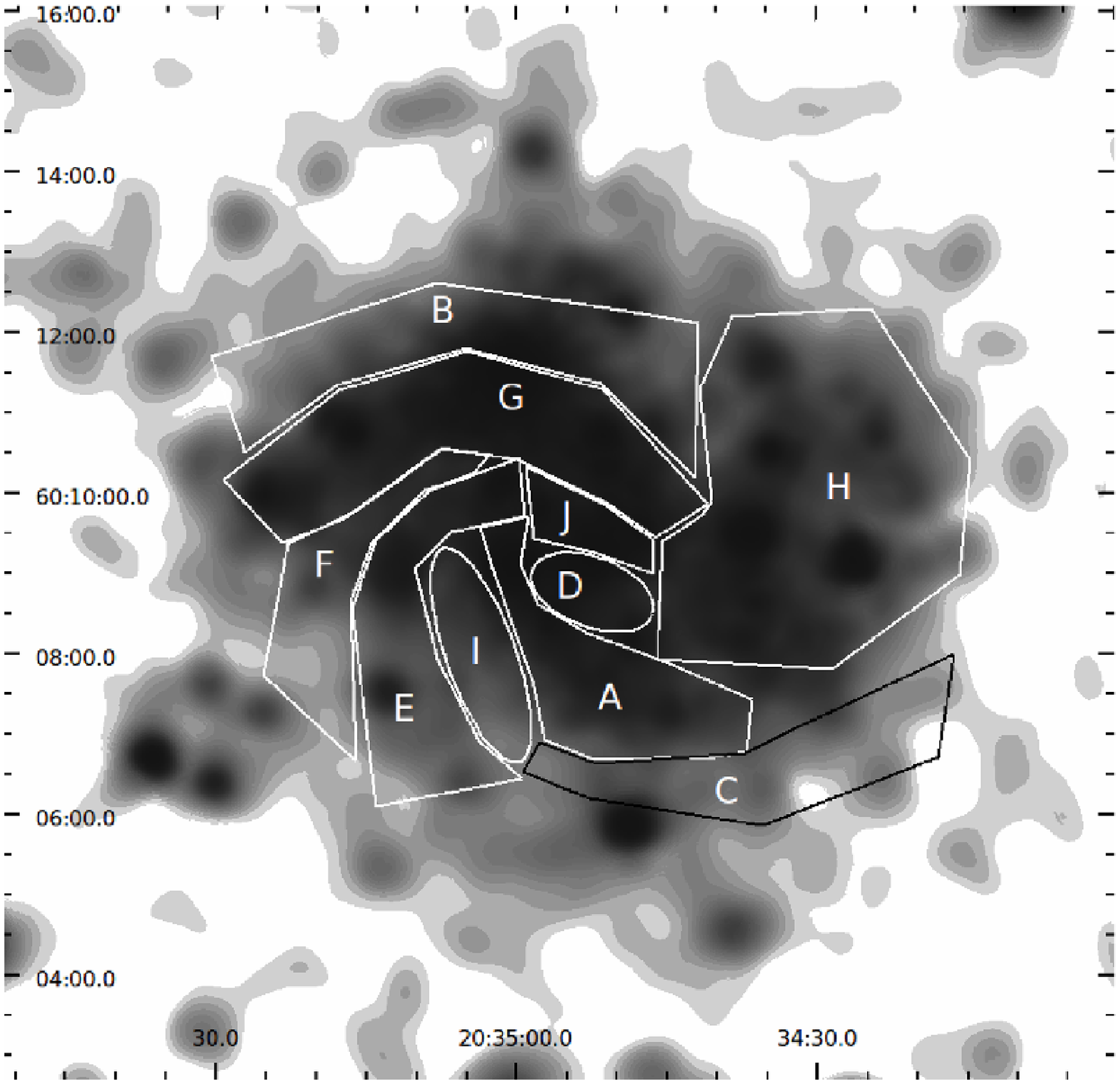}}
\end{center}
\caption{
         Regions of diffuse X-ray emission from NGC\,6946 (see text for a detailed description) overlaid on the same maps as in
Fig~\ref{6946points}.
        }
\label{6946xregs}
\end{figure*}

\begin{figure}[ht]
                        \resizebox{\hsize}{!}{\includegraphics[clip]{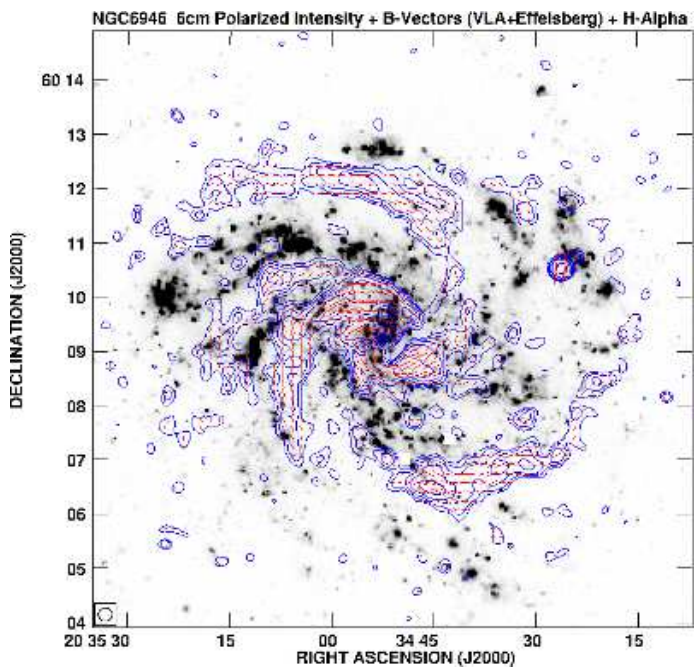}}
                \caption{Map of radio-polarized intensity at 4.85\,GHz ($\lambda$ 6\,cm) of NGC\,6946 \citep[from][]{beck07}. The red lines show the orientation
                        of the magnetic fields.}
                \label{6946radio}
\end{figure}

\begin{figure*}[ht]
\begin{center}
\resizebox{0.45\hsize}{!}{\includegraphics[angle=-90]{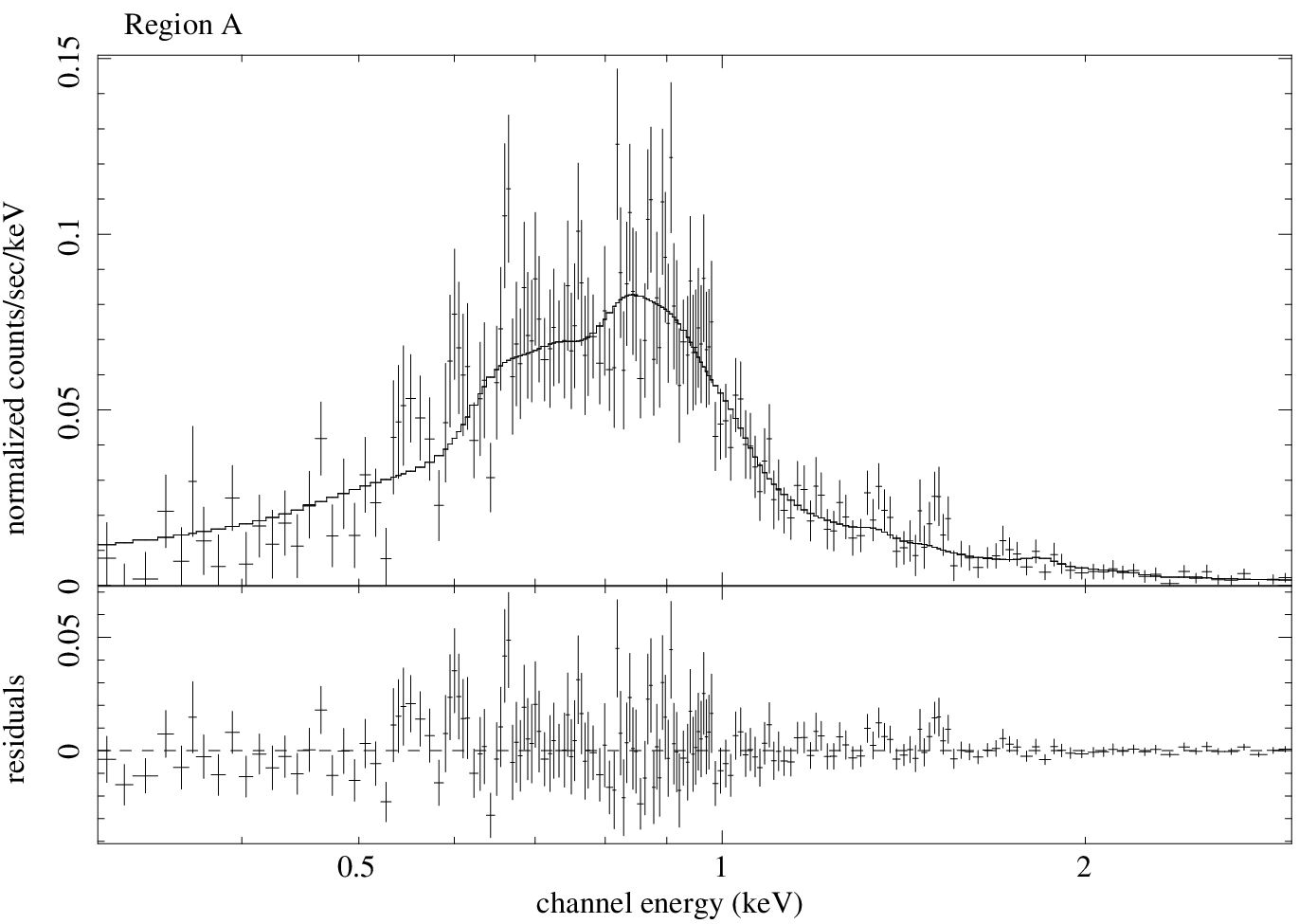}}
\resizebox{0.45\hsize}{!}{\includegraphics[angle=-90]{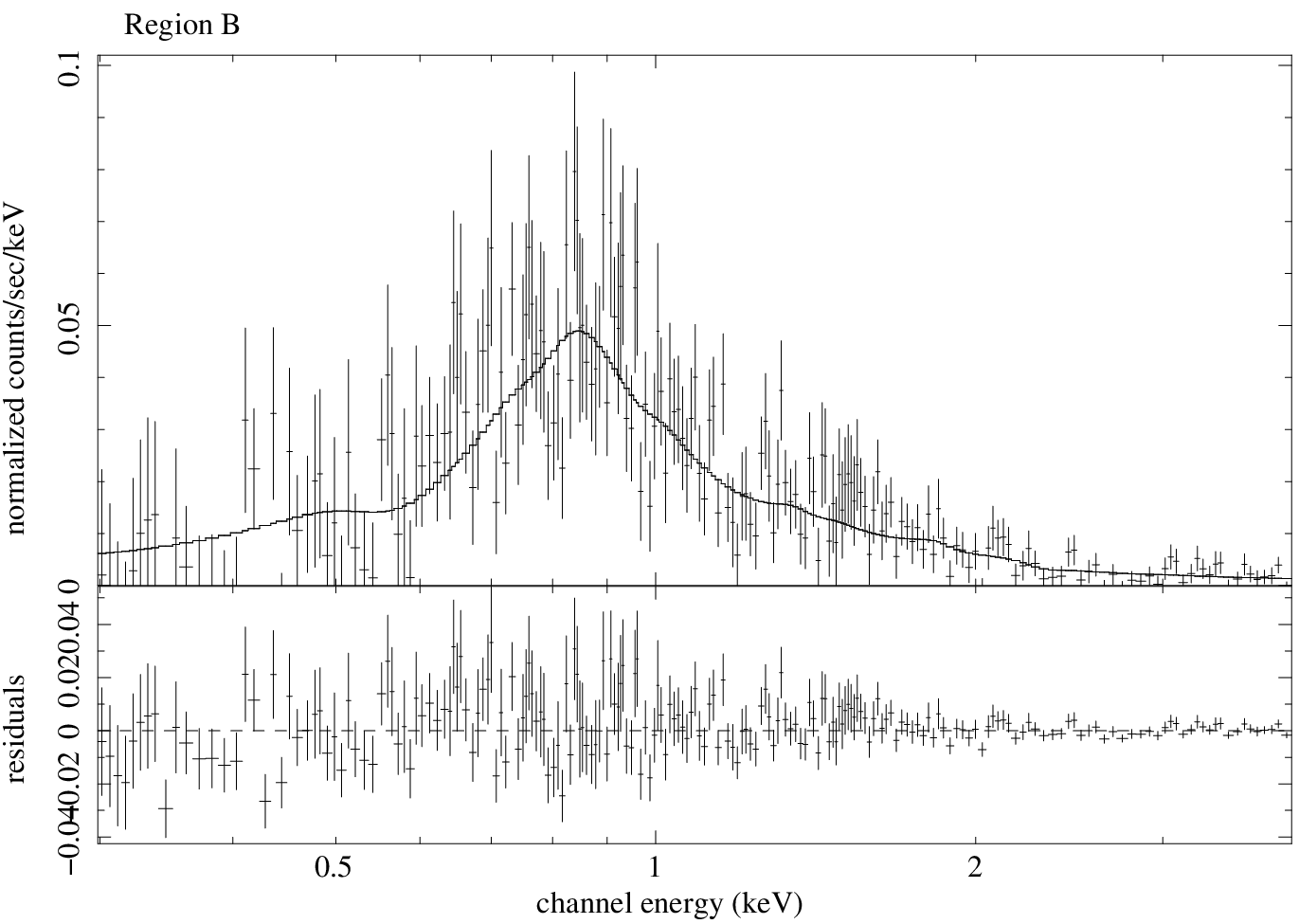}}
\resizebox{0.45\hsize}{!}{\includegraphics[angle=-90]{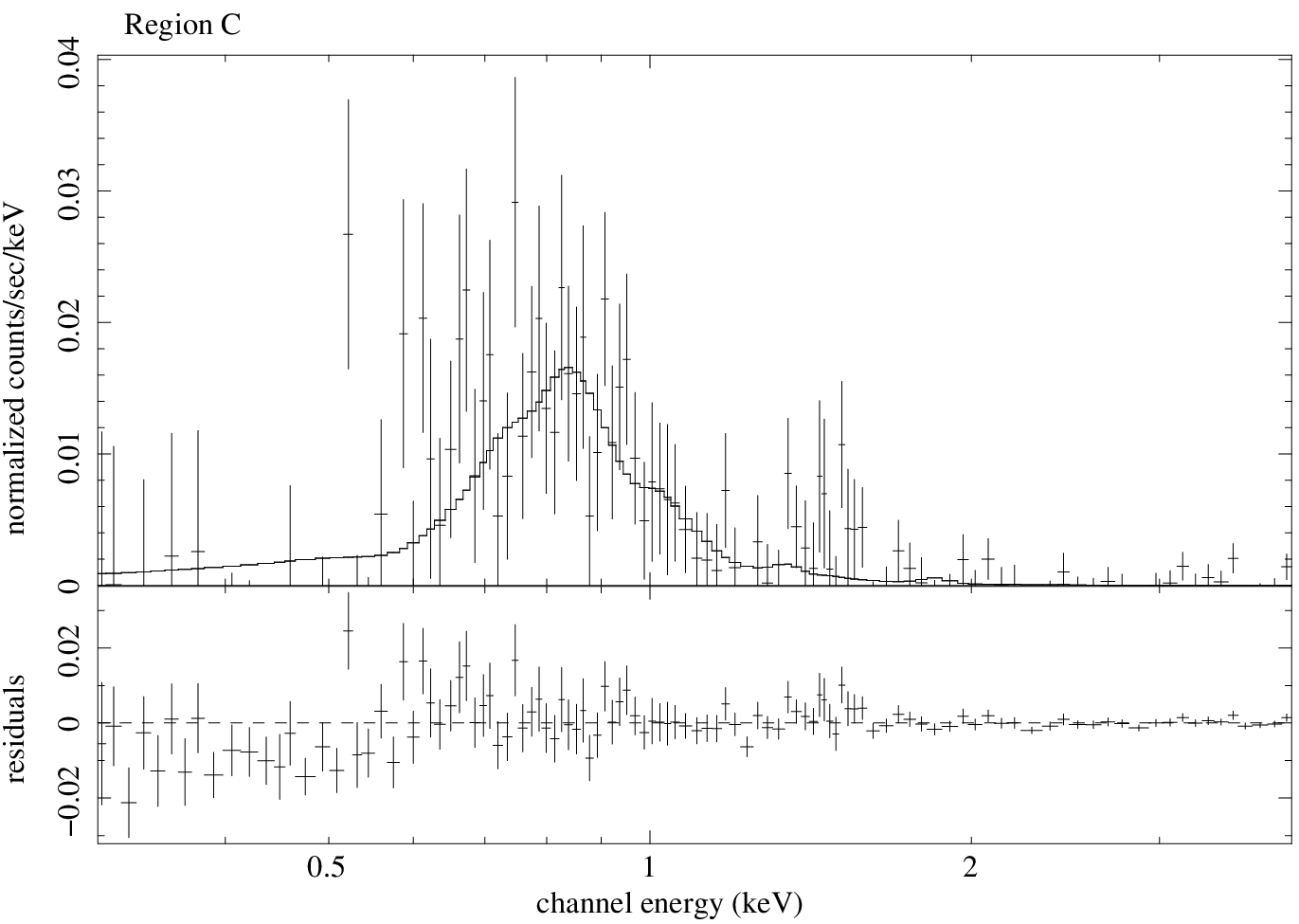}}
\resizebox{0.45\hsize}{!}{\includegraphics[angle=-90]{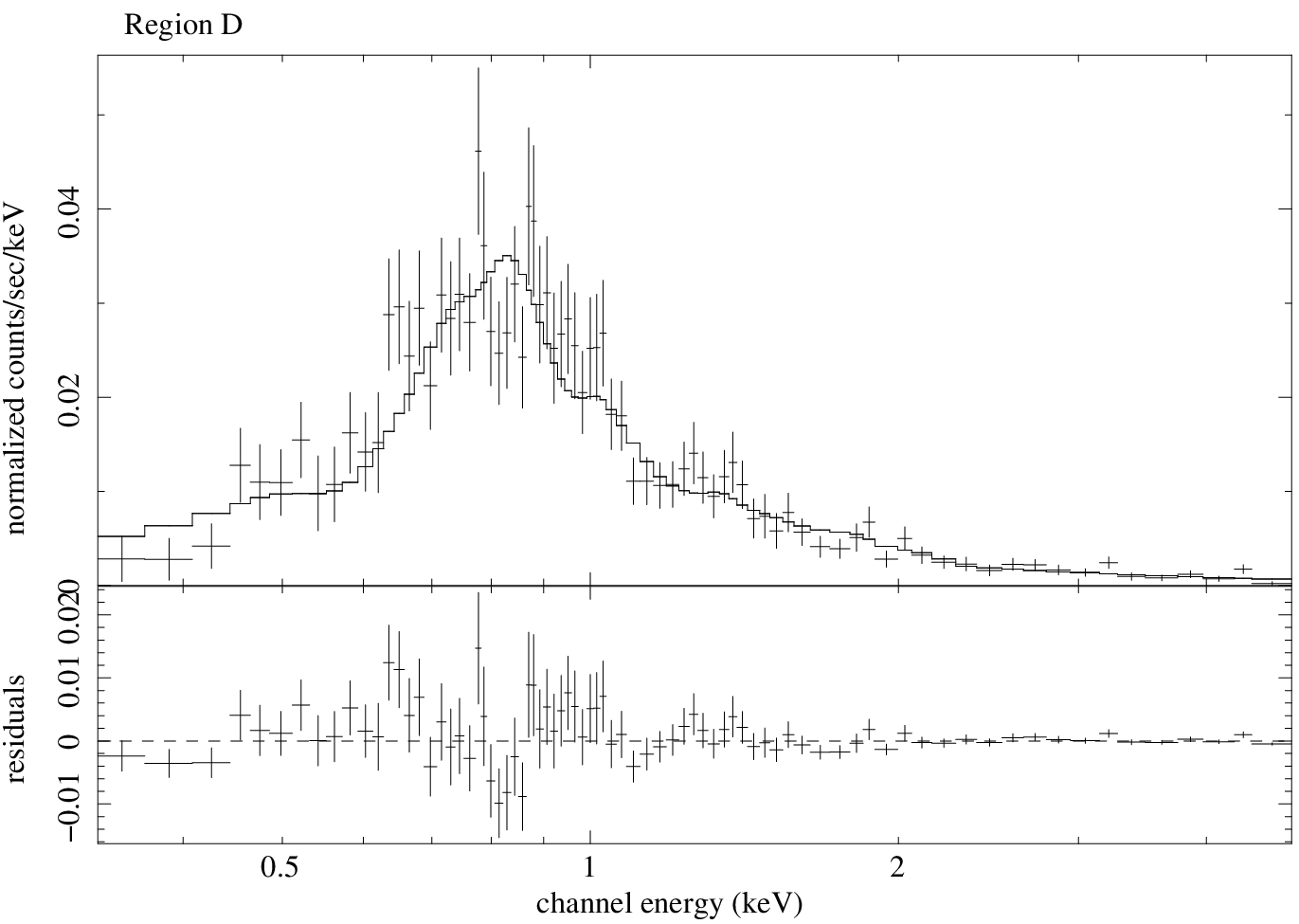}}
\resizebox{0.45\hsize}{!}{\includegraphics[angle=-90]{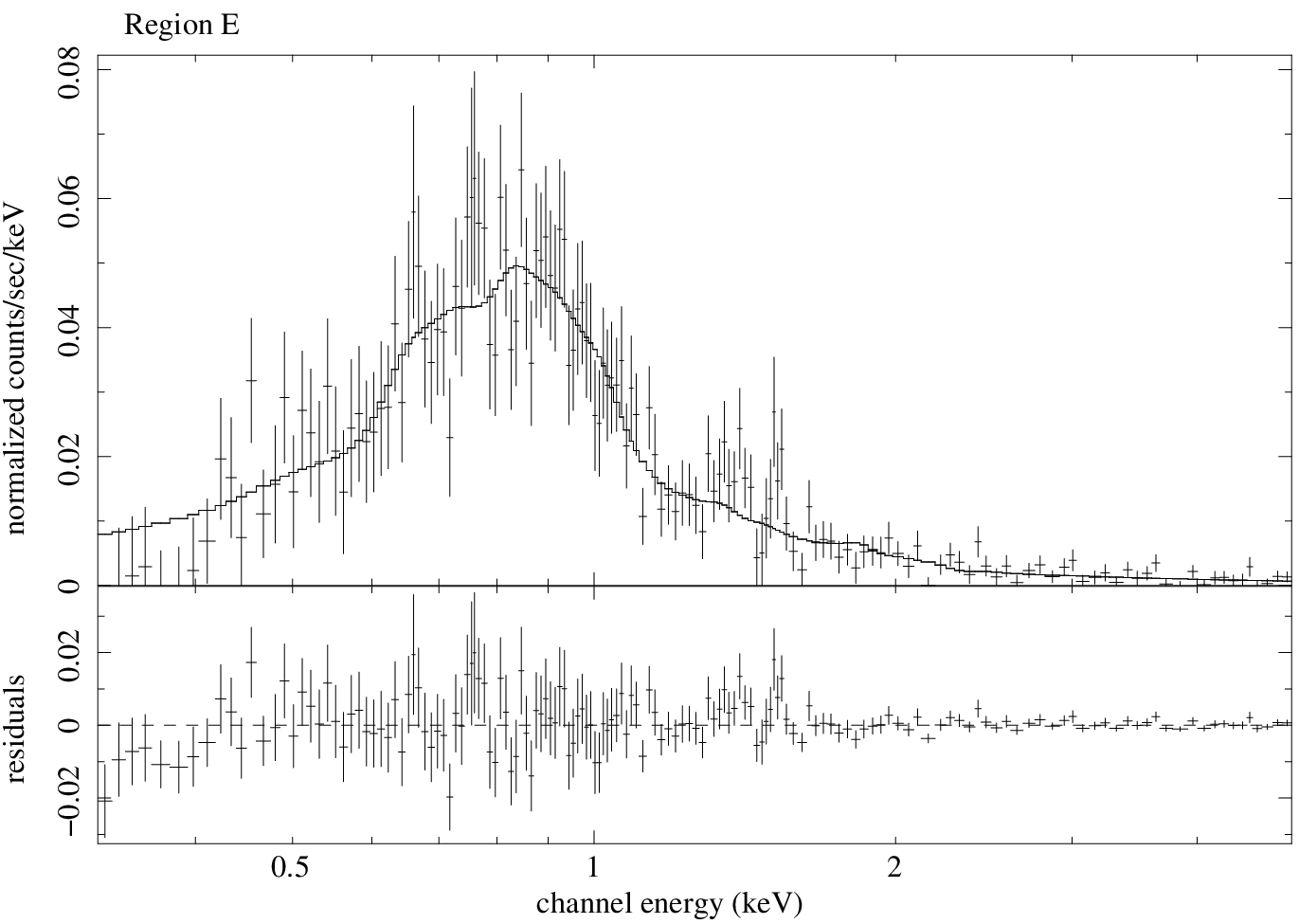}}
\resizebox{0.45\hsize}{!}{\includegraphics[angle=-90]{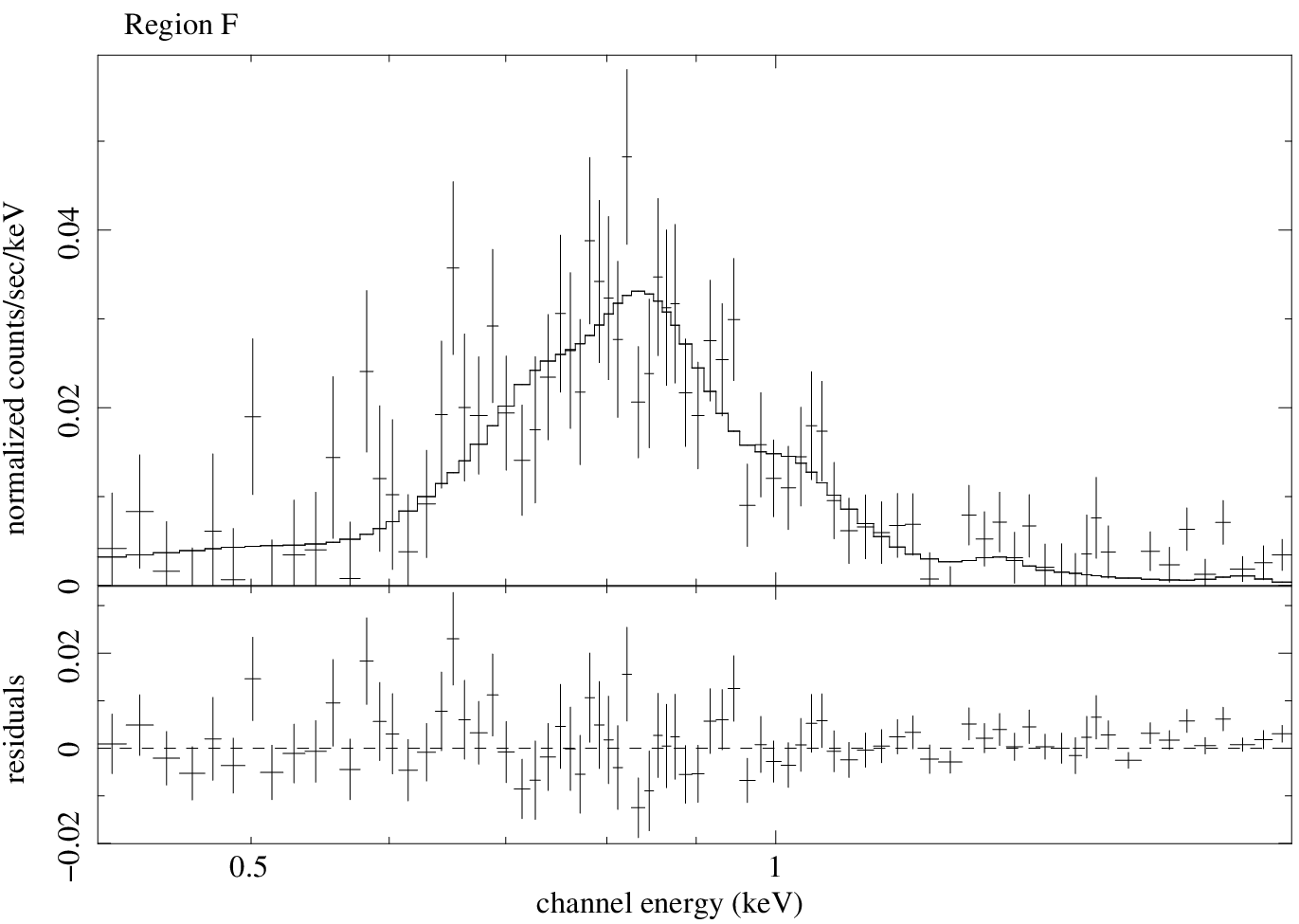}}
\resizebox{0.45\hsize}{!}{\includegraphics[angle=-90]{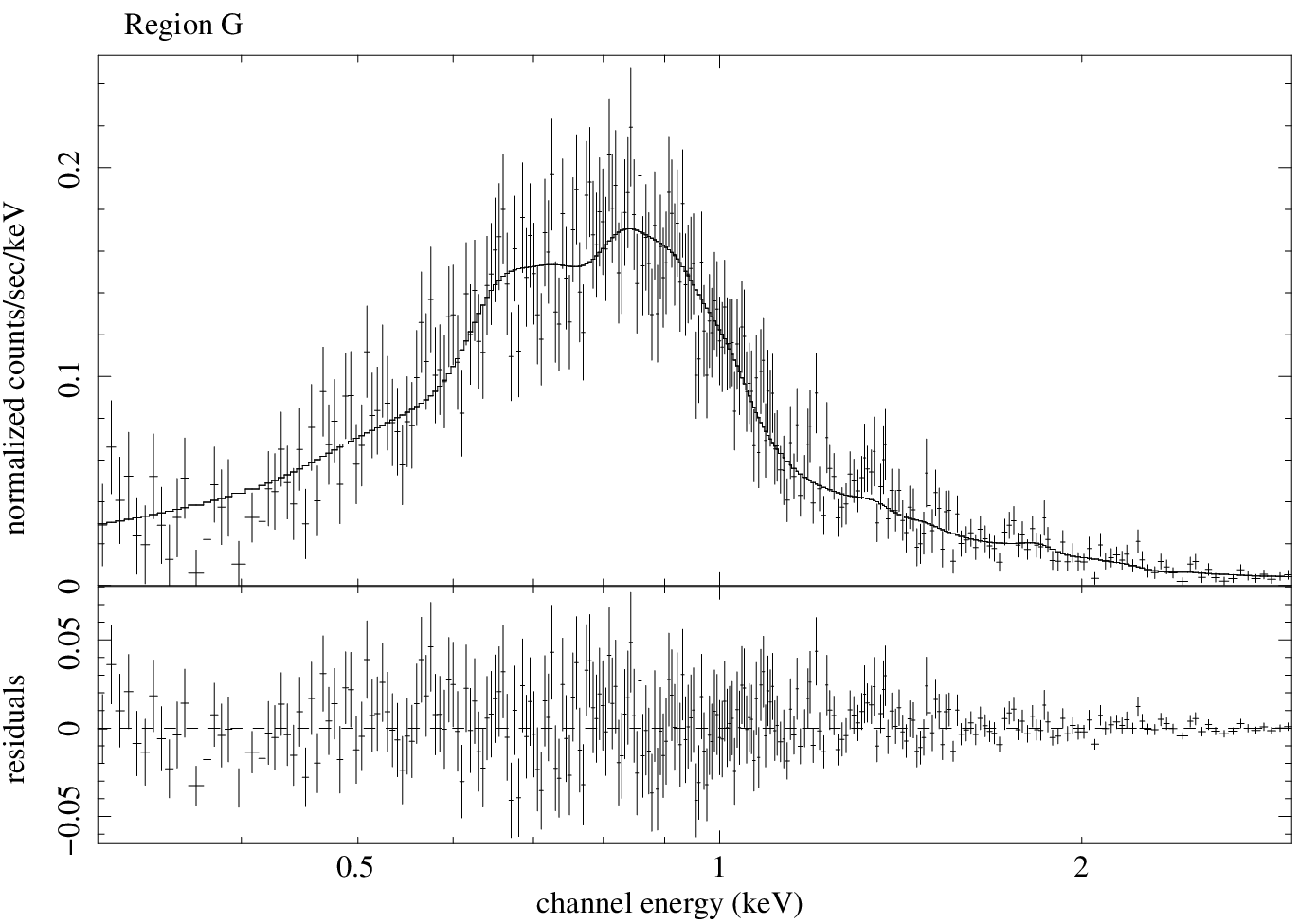}}
\resizebox{0.45\hsize}{!}{\includegraphics[angle=-90]{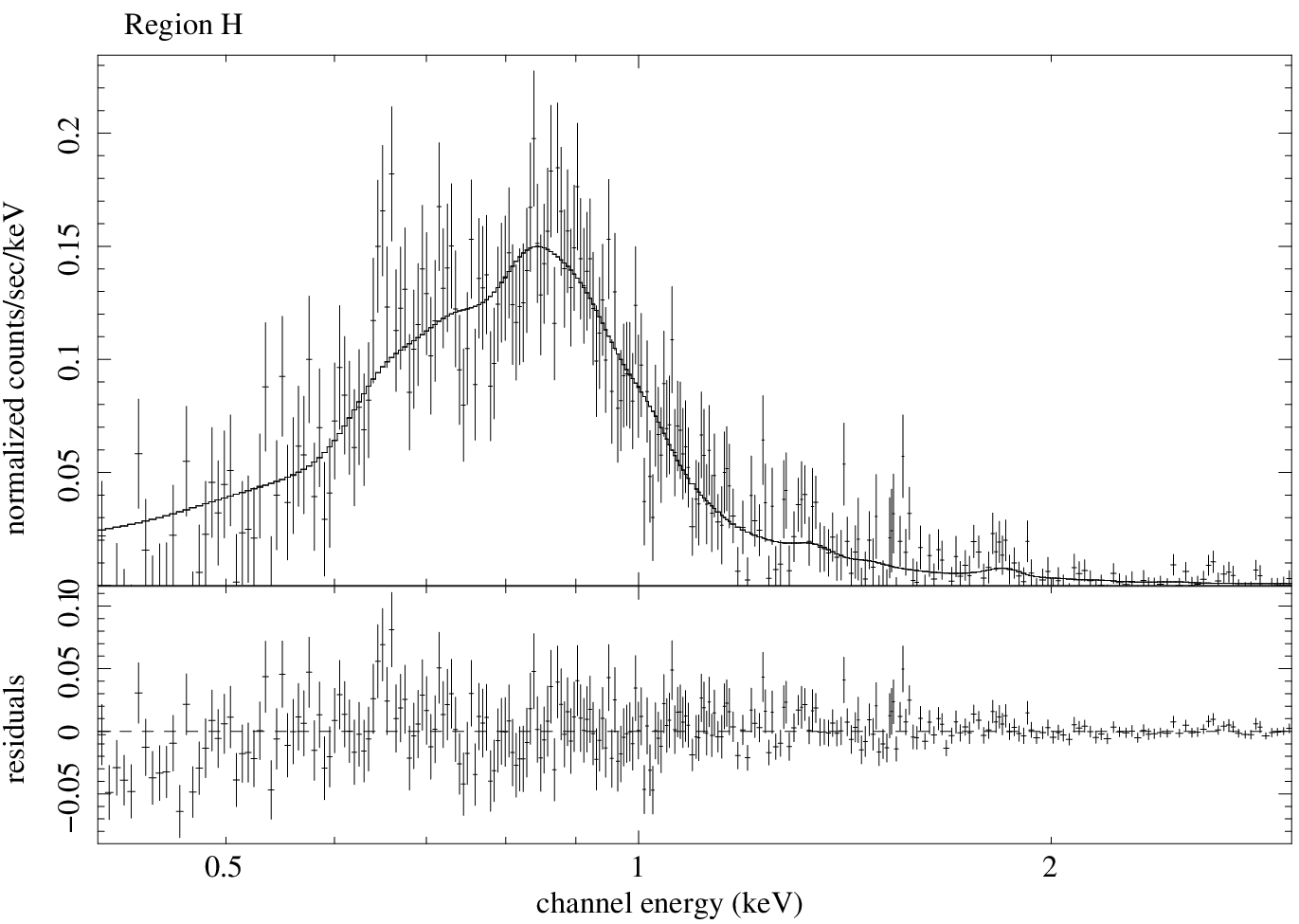}}
\end{center}
\caption{Model fits to the regions of diffuse X-ray emission in NGC\,6946. See Tables~\ref{6946xtabr} and
\ref{6946xfr}.}
\label{6946mod1}
\end{figure*}

\begin{figure*}[ht]
\begin{center}
\resizebox{0.45\hsize}{!}{\includegraphics[angle=-90]{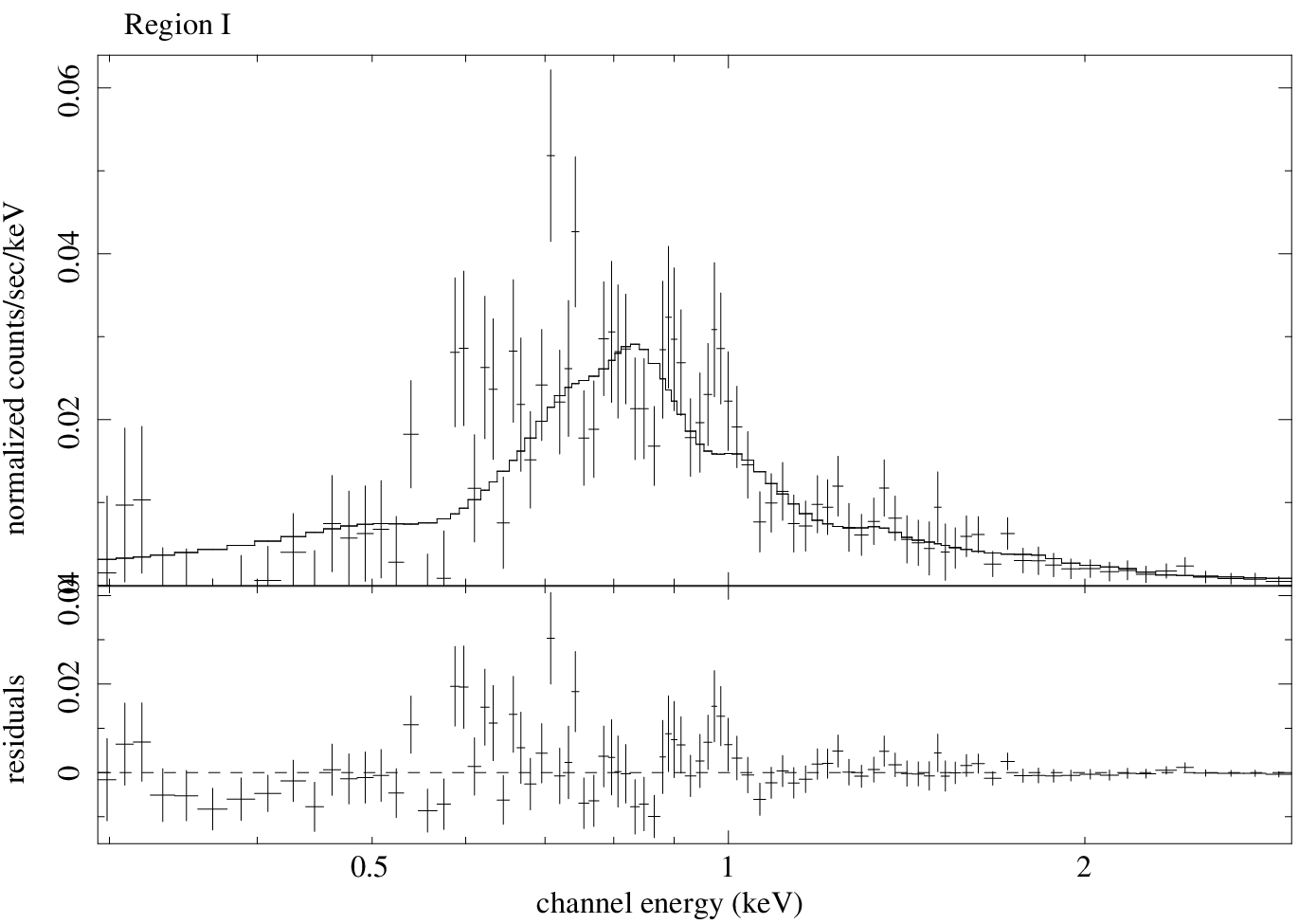}}
\resizebox{0.45\hsize}{!}{\includegraphics[angle=-90]{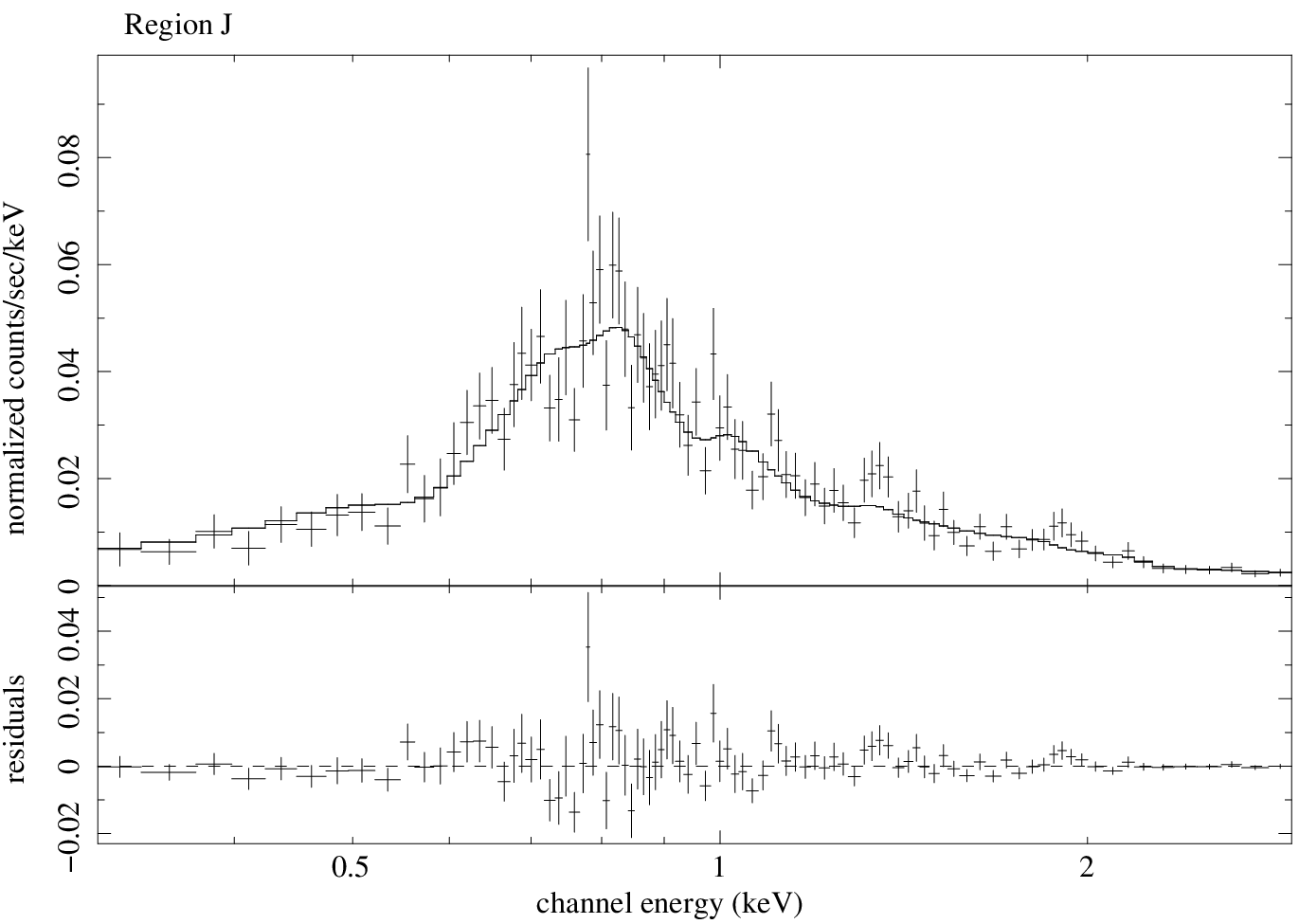}}
\end{center}
\caption{Model fits to the regions of diffuse X-ray emission in NGC\,6946. See Tables~\ref{6946xtabr} and
\ref{6946xfr}.}
\label{6946mod2}
\end{figure*}

To analyse emission of the hot gas from NGC\,6946, we used H$\alpha$ and UV images to choose areas that correspond to the star-forming regions. The radio morphology of the galaxy was
also taken into account because the most prominent polarized features were found {\em \textup{between}} the gaseous spiral arms of NGC\,6946 (see Fig.~\ref{6946radio}).
All selected regions are presented in Fig.~\ref{6946xregs}. A brief description of all regions is presented in Table~\ref{names}.
For transparency, region letters are used throughout this paper.
In the process of extracting the spectra, the emission from point sources (Fig.~\ref{6946points}) was excluded.

\begin{table*}[ht]
\caption{\label{names}Regions in NGC\,6946 used for the spectral analysis.}
\centering
\begin{tabular}{cl}
\hline\hline
Region                  & Region                        \\
letter                  & description                   \\
\hline
\vspace{5pt}
A               & south-western arm  \\
\vspace{5pt}
B               & north-western interarm  \\
\vspace{5pt}
C               & south-western interarm  \\
\vspace{5pt}
D               & central region w/o UV emission  \\
\vspace{5pt}
E               & south-eastern arm  \\
\vspace{5pt}
F               & eastern interarm  \\
\vspace{5pt}
G               & northeastern arm  \\
\vspace{5pt}
H               & western arms  \\
\vspace{5pt}
I               & south-eastern interarm  \\
\vspace{5pt}
J               & central region with UV emission \\
\hline
\end{tabular}
\end{table*}

To investigate the temperature of the hot gas in selected regions of NGC\,6946, we fitted a single thermal plasma model, adding a power-law component to account
for undetected point sources and/or residual emission from excluded sources. Only regions C and F did not require this additional power-law component.
This could simply be caused by a lower number of net counts in their spectra, resulting in a low signal-to-noise ratio, hence lower accuracy of the fitting, 
so that the basic models are equally good in fitting the data. 
On the other hand, both spectra show very little emission above 1\,keV, which suggests that the harder emission from point sources
contributes hardly anything.

In Table~\ref{singles} we present the results of single thermal plasma model fitting. For regions of the spiral galactic arms (regions A, E, G, and H),
very steep photon indices of the power-law component are visible, which is unexpected for typical galactic X-ray point sources.
This might be due to enhanced emission in the soft-to-medium energy band (around and above 1\,keV), however, which can also suggest
that a second thermal component is necessary to account for the significant emission from the hot gas in the galactic disk. To check this possibility, we fitted
a model consisting of two thermal components for the areas of the spiral arms and kept the previously introduced power-law component. Since both parameters and residuals of the new fits
were much more physical, that is, showed values in expected ranges, we used them as the final fits for the subsequent analysis. 

To ensure that a single thermal plasma model is the best-fit model for the remaining regions (central areas and interarm regions), we also fitted a model with two thermal 
components to their corresponding spectra. In all cases we obtained significantly flatter (lower) photon indices of the power-law component, which suggests 
that some of the harder emission from the unresolved point sources was fitted with the new second thermal component. 
Furthermore, most of the fitted parameters were poorly constrained. For some of them it was impossible to get any constraints.
The final models for all regions are presented in Table~\ref{6946xtabr}.

For regions of the spiral arms for which a model with two thermal components was used, we assumed that the cooler component ($\sim$ 0.3\,keV) of the hot gas is associated with the galactic halo
and the hotter component ($\sim$ 0.7\,keV) with the emission from the disk, as is observed in edge-on galaxies \citep[e.g.][]{tuellmann06}.

For the remaining regions, that is, for the central areas and the interarm regions, we assumed that the single thermal component can be described as a mixture 
of gas components from the disk and the halo above. The parameters of these components seem to be similar to the level, which does not allow clearly separating them at a given sensitivity level. For the interarm regions this can be easily explained by the lack of star-forming regions in the disk, 
which causes the gas to be relatively uniform throughout the entire volume. Nevertheless, we cannot use this argument for the central regions of the galaxy, 
where the disk emission is certainly significant. A reasonable explanation for this case is a high star-forming activity of NGC\,6946 and the consequent 
internal absorption that is highest in the central densest part of the galactic disk. This absorption may cause some of the disk emission from the 
central regions I and J to remain hidden, and the rest, together with the halo gas, mimics a homogenous medium.        

\begin{table*}[ht]
\caption{\label{singles}Parameters of single plasma model-fits to the diffuse emission regions studied in NGC\,6946.}
\centering
\begin{tabular}{cccc}
\hline\hline
Region          & kT                            &       Photon                  & Reduced \\
\vspace{5pt}    & [keV]                         &       index                   & $\chi^2$\\
\hline
\vspace{5pt}
A               & 0.53$^{+0.05}_{-0.08}$        & 2.55$^{+0.26}_{-0.17}$        & 1.21  \\
\vspace{5pt}
B               & 0.60$^{+0.06}_{-0.08}$        & 1.84$\pm$0.22                 & 1.13  \\
\vspace{5pt}
C               & 0.55$^{+0.10}_{-0.11}$        & --                            & 1.15  \\
\vspace{5pt}
D               & 0.49$^{+0.07}_{-0.14}$        & 1.78$^{+0.13}_{-0.08}$        & 0.99  \\
\vspace{5pt}
E               & 0.53$^{+0.04}_{-0.12}$        & 2.31$^{+0.20}_{-0.23}$        & 1.17  \\
\vspace{5pt}
F               & 0.53$\pm$0.06                 & --                            & 1.05  \\
\vspace{5pt}
G               & 0.50$^{+0.02}_{-0.03}$        & 2.54$^{+0.05}_{-0.11}$        & 1.35  \\
\vspace{5pt}
H               & 0.53$^{+0.08}_{-0.03}$        & 2.68$^{+0.45}_{-0.40}$        & 1.24  \\
\vspace{5pt}
I               & 0.50$^{+0.14}_{-0.13}$        & 1.86$^{+0.29}_{-0.25}$        & 1.07  \\
\vspace{5pt}
J               & 0.42$^{+0.09}_{-0.05}$        & 1.69$^{+0.19}_{-0.16}$        & 1.01  \\
\hline
\end{tabular}
\end{table*}

\begin{table*}[ht]
\caption{\label{6946xtabr}Final model-fit parameters for the regions studied in NGC\,6946.}
\centering
\begin{tabular}{clcccrr}
\hline\hline
Region          & Model                         & kT$_1$                     & kT$_2$                        &       Photon                  &$\chi_{\rm red}^2$& Net   \\
                & type                          & [keV]                      & [keV]                         &       Index                   &                & counts\\
\hline
\vspace{5pt}
A               & wabs(mekal+mekal+power law)    & 0.28$^{+0.03}_{-0.04}$        & 0.73$^{+0.07}_{-0.06}$  & 1.82$^{+0.36}_{-0.44}$        &       0.96       & 3512        \\
\vspace{5pt}
B               & wabs(mekal+power law)          & 0.60$^{+0.06}_{-0.08}$        & --                              & 1.84$\pm$0.22                 &       1.13       & 2516        \\
\vspace{5pt}
C               & wabs*mekal                    & 0.55$^{+0.10}_{-0.11}$        & --                              & --                            &       1.15       &  385        \\
\vspace{5pt}
D               & wabs(mekal+power law)          & 0.49$^{+0.07}_{-0.14}$        & --                             & 1.78$^{+0.13}_{-0.08}$        &       0.99       & 1581        \\
\vspace{5pt}
E               & wabs(mekal+mekal+power law)    & 0.30$^{+0.05}_{-0.03}$        & 0.80$^{+0.09}_{-0.14}$  & 1.72$^{+0.26}_{-0.21}$        &       0.99       & 2552        \\
\vspace{5pt}
F               & wabs*mekal                    & 0.53$\pm$0.06                 & --                              & --                            &       1.05       &  896        \\
\vspace{5pt}
G               & wabs(mekal+mekal+power law)    & 0.27$\pm$0.02                 & 0.78$\pm$0.06                   & 1.93$^{+0.17}_{-0.19}$        &       1.03       & 8179        \\
\vspace{5pt}
H               & wabs(mekal+mekal+power law)    & 0.28$^{+0.04}_{-0.03}$        & 0.68$^{+0.07}_{-0.05}$  & 1.59$^{+0.70}_{-0.88}$        &       1.09       & 5008        \\
\vspace{5pt}    
I               & wabs(mekal+power law)          & 0.50$^{+0.14}_{-0.13}$        & --                             & 1.86$^{+0.29}_{-0.25}$        &       1.07       & 1211        \\
\vspace{5pt}
J               & wabs(mekal+power law)          & 0.42$^{+0.09}_{-0.05}$        & --                              & 1.69$^{+0.19}_{-0.16}$        &       1.01       & 2349        \\
\hline
\end{tabular}
\tablefoot{
\tablefoottext{a}{Column density in [10$^{21}$ cm$^{-2}$].}
}
\end{table*}

Region B shows almost no H$\alpha$ or UV emission, while region J includes the central parts of the galaxy. Both regions differ significantly in the value of the fitted temperature
of the hot gas. It is much hotter (0.60$^{+0.06}_{-0.08}$\,keV) for the quiet region B than for region J with its clear galactic-disk (nuclear) emission (0.42$^{+0.09}_{-0.05}$\,keV).
The remaining single thermal component model fits show that the
temperature of the hot gas is between 0.5\,keV and 0.55\,keV. Halo components of the model fits to the regions of spiral
arms show a very constant temperature of around 0.28\,keV. For the disk components, however, differences are observed, 
as the south-western and western part of the disk (spiral arms A and H) have a significantly lower temperature than the eastern and north-eastern part (spiral arms E and G).

For all regions except for region H (western spiral arm) most of the X-ray flux is produced by the power-law component that
is attributed to unresolved and/or residual emission from point sources.
For the exceptional region H we note, however, that the value of the power-law flux is only poorly constrained. Therefore, its value might be much higher, which would lead to
a higher contribution to the total flux. A high contribution from the power-law components of the fits might arise because NGC\,6946 is a starburst galaxy with
a large population of X-ray point sources (Fig.~\ref{6946points}). The method of source extraction used in this paper, described in Sect.~\ref{obsred}, might certainly
cause this effect. However, as argued before, this ensured that as little as possible of the diffuse emission has been lost by point-source extraction.

For the regions of the spiral arms, more flux comes from the halo components according to the fits, which agrees with a physical picture where halo emission is produced
in a much larger volume density of the underlying star formation.
These fractions are
equal only for the western spiral arm (region H); we discuss
this in Sect.~\ref{gasparams} in more detail.

\begin{table*}[ht]
        \caption{\label{6946xfr}Total (0.3 - 12 keV) unabsorbed fluxes in 10$^{-14}$erg\,cm$^{-2}$s$^{-1}$ for modelled regions in NGC\,6946.}
\centering
\begin{tabular}{ccccc}
\hline\hline
\vspace{5pt} Region     & mekal cold                    & mekal hot                     & power law                      & total                         \\
\hline
A                       & 3.5$^{+0.8}_{-0.9}$  (0.32)   & 2.5$^{+0.8}_{-0.6}$  (0.23) & 4.8$^{+5.1}_{-2.0}$  (0.45)   & 10.8$^{+6.7}_{-3.5}$          \\
\vspace{5pt}
B                       & 2.4$^{+0.3}_{-0.5}$  (0.22)   & --                            & 8.5$^{+4.0}_{-2.5}$  (0.78)     & 10.9$^{+4.4}_{-2.9}$          \\
\vspace{5pt}
C                       & 1.1$^{+0.1}_{-0.2}$  (1.00)   & --                            & --                              & 1.1$^{+0.1}_{-0.2}$           \\
\vspace{5pt}
D                       & 1.4$^{+0.2}_{-0.3}$  (0.25)   & --                            & 4.2$^{+1.0}_{-1.1}$  (0.75)     & 5.6$^{+1.2}_{-1.4}$           \\
\vspace{5pt}
E                       & 2.6$\pm$0.9          (0.26)   & 1.6$^{+0.7}_{-0.6}$  (0.16) & 5.7$^{+3.5}_{-2.2}$  (0.58)   & 9.8$^{+5.0}_{-3.74}$          \\
\vspace{5pt}
F                       & 2.1$^{+0.1}_{-0.2}$  (1.00)   & --                            & --                              & 2.1$^{+0.1}_{-0.2}$           \\
\vspace{5pt}
G                       & 9.3$^{+1.5}_{-1.4}$  (0.31)   & 5.2$^{+1.0}_{-1.3}$  (0.18) & 15.3$^{+5.4}_{-3.3}$ (0.51)   & 29.7$^{+7.8}_{-6.0}$          \\
\vspace{5pt}
H                       & 6.6$^{+1.9}_{-1.7}$  (0.43)   & 6.4$^{+1.5}_{-1.8}$  (0.42) & 2.3$^{+12.8}_{-1.9}$ (0.15)   & 15.4$^{+16.0}_{-5.5}$         \\
\vspace{5pt}
I                       & 1.2$\pm$0.2          (0.30)   & --                            & 2.8$^{+1.7}_{-1.1}$  (0.70)     & 4.0$^{+1.8}_{-1.3}$           \\
\vspace{5pt}
J                       & 2.0$\pm$0.5          (0.21)   & --                            & 7.5$^{+2.6}_{-1.8}$  (0.79)     & 9.5$^{+3.1}_{-2.3}$           \\
\hline
\end{tabular}
\tablefoot{
Values in parentheses are fractions of the total flux for a given component.
}
\end{table*}

\FloatBarrier

\section{Discussion}
\label{disc}

\subsection{Point sources}
\label{pts}

Of the 35 point sources found in the galactic disk of NGC\,6946, for 19 of them we were able to calculate only the hardness ratios. All values
are presented in Table~\ref{6946sources}. For the remaining sources we fitted models to the acquired spectra (see Sect.~\ref{sourcespec}).
As mentioned above, all sources except for source 27 required an additional thermal component in the model. This was most likely due to the characteristics
of the observations with the XMM-Newton telescope, that is, a
relatively low resolution with a high sensitivity to diffuse X-ray emission. As a result, all spectra
of the studied point sources included significant information from the diffuse gas present across the entire disk of NGC\,6946.

For several sources, an additional component to account for internal absorption was needed to obtain a good fit (see Table~\ref{6946xtabs}).
For three sources, 19, 23, and 32, this was crucial. For source 19, which is the core region of NGC\,6946 and therefore not just a point source,
this additional absorption is easy to explain because we observe the densest parts of the galaxy.
The two other sources, however, are located in areas relatively
free of star formation (see Fig.~\ref{6946points}, left). Furthermore, none
of the extended regions used for the spectral analysis of the diffuse emission coming from the hot gas needed an additional internal absorption component to
be described by the model. Therefore, the argument of (relatively) low-resolution observations of a gas- and dust-rich galaxy cannot be used here.
Instead, for each source that required an internal absorption component in the model, we propose that it is an X-ray binary
(or an intermediate-mass black hole, IMBH) surrounded by a dust torus.

To calculate luminosities of the spectrally analysed point sources, we only used the fluxes of the power-law component to exclude the contribution from the
galactic hot gas in the modelled spectrum (Table~\ref{6946xfs}). Half of the sources show luminosities higher than 10$^{39}$erg\,s$^{-1}$. This value is
the most widely used observational definition of an ultra-luminous source (ULX); it is often associated with an accreting IMBH that
forms in the core
collapse of young dense stellar clusters \citep[e.g.][]{miller}. Since the abundance of ULXs is often linked to recent star formation activity
\citep[e.g.][]{berghea13}, a significant number of ULXs in NGC\,6946 agrees well with the vivid star formation of this galaxy.

One of these sources is the nebula MF\,16 (region 26) mentioned
above.
Although some of its thermal emission (from both model components) can be associated with the galactic hot gas, as mentioned above,
a significant contribution from the hotter component needs a different explanation. That the gas temperature exceeds 1\,keV (Table~\ref{6946xtabs}),
which is much higher than the temperature of the disk gas,
may provide further evidence that MF16 might be indeed a supernova remnant, as previously claimed by \citet{matonick}, and the very high temperature of the hot gas
may result from shock heating of the medium surrounding the explosion region. This contradicts the findings of \citet{berghea}, who claimed that no
signs of shock heating are present.

\subsection{Extended emission}
\label{diff}

\subsubsection{Parameters of the hot gas}
\label{gasparams}

From our spectral model fits we were able to derive more parameters of the hot gas, including electron densities $n_e$, masses $M$, thermal energies $\epsilon_{th}$, and
cooling times $\tau$. To perform our calculations we used the model of thermal cooling and ionization equilibrium of \citet{Nulsen}, where $L_X=1.11\cdot \Lambda(T)\,n^2_e\,V\,\eta$,
$\eta$ is an unknown filling factor and $\Lambda(T)$ is a cooling coefficient of the order of $10^{-22}\,{\rm erg}\,{\rm cm}^3\,s^{-1}$ for temperatures of a few millions K
\citep{Raymond}. However, the main difficulty in calculating
the physical parameters of the hot gas component are the assumptions about the emitting volume $V$.

It seems straightforward to assume that we only see soft X-ray emission from the visible side of the disk (and therefore the halo),
with all emission from the other side being absorbed by the neutral
hydrogen in the galactic disk. For the visible part of the halo emission, we assumed a cylindrical volume above the disk of NGC\,6946, extending out to half of the $D_{25}$
diameter of the galaxy (10\,kpc). In this geometry, the halo around NGC\,6946 would be roughly spherical. This approach seems to be justified for a starbust galaxy that has significantly extended halos of the X-ray emission. For the disk emission we assumed a disk thickness of 1\,kpc. For the interarm regions, for which a single thermal 
model was fitted, we used volumes of 10\,kpc times the area of a region, as the model describes the mixed emission from both the disk and the halo (see Sect.~\ref{regspec}). 

Tables~\ref{halogas} and \ref{diskgas} present the derived parameters of the hot gas in the areas of diffuse emission in NGC\,6946.

To verify the obtained values, a comparison with earlier Chandra observations of NGC\,6946 by \citet{schlegel03} would be an important step. Because of the low sensitivity to diffuse emission of these observations, the analysis was unfortunately made only for the entire galactic disk and no detailed study is available. 
Still, their temperatures of 0.25$\pm$0.03\,keV and 0.70$\pm$0.10\,keV for two thermal components agree very well with our values for the halo and the 
disk gas temperatures, respectively. The electron density of $\sim$0.012 ($\times\,\eta^{-0.5}$)\,cm$^{-3}$ derived by \citet{schlegel03} for the 1\,kpc thick disk also 
matches our results well if we consider a ten times larger volume than used for our calculations. A similar study of six other nearby face-on late-type spiral galaxies 
was reported by \citet{owen}, who obtained temperatures of two thermal model fits of 0.2-0.3\,keV and 0.6-0.7\,keV and the derived electron densities of the gas of the order of 
a few $10^{-3}$ ($\times\,\eta^{-0.5}$)\,cm$^{-3}$, depending on the level of the star-forming activity.

\subsubsection{Hot gas components}
\label{components}

As mentioned before, for the regions of the spiral arms we needed a two-temperature model to account for the emission from the hot gas residing
in both galactic disk and the surrounding halo.
Although the temperatures of the hot gas are almost identical for the halo components and similar
for the disk components (where two pairs could be distinguished - A with H and E with G - see Table~\ref{6946xtabr}),
the derived parameters show significant differences (see Table~\ref{diskgas}). As expected, for the north-eastern spiral arm with the brightest H$\alpha$ and UV emission,
marked as region G, we obtained the highest values of number density and energy density of all spiral arms.

\begin{table*}[ht]
\caption{\label{halogas}Derived parameters of the hot gas in the interarm and central regions of NGC\,6946.}
\centering
\begin{tabular}{cccccc}
\hline\hline
Region & (n$_{halo}\eta^{-0.5}$)\tablefootmark{a} & (M$^{halo}_{gas}\eta^{0.5}$)\tablefootmark{a}& (E$^{halo}_{th}\eta^{0.5}$)\tablefootmark{a}& ($\epsilon^{halo}_{th}\eta^{-0.5}$)\tablefootmark{a} & ($\tau^{halo}\eta^{0.5}$)\tablefootmark{a} \\
\vspace{5pt}&[10$^{-3}$cm$^{-3}$]                 & [10$^6$M$_\odot$]                            & [10$^{54}$\,erg]                                     & [10$^{-12}$\,erg\,cm$^{-3}$]   & [Myr]          \\
\hline
\vspace{5pt}
B                       & 0.84$\pm$0.07                 & 5.09$^{+0.37}_{-0.47}$        & 8.74$^{+1.58}_{-1.87}$  & 1.22$^{+0.22}_{-0.26}$  & 1969$^{+98}_{-13}$           \\
\vspace{5pt}
C                       & 0.72$^{+0.07}_{-0.06}$        & 2.66$^{+0.26}_{-0.24}$        & 4.20$^{+1.24}_{-1.14}$  & 0.96$^{+0.28}_{-0.26}$  & 2063$^{+388}_{-277}$        \\
\vspace{5pt}
D                       & 2.03$^{+0.15}_{-0.12}$        & 1.22$^{+0.09}_{-0.07}$        & 1.71$^{+0.39}_{-0.56}$  & 2.40$^{+0.54}_{-0.79}$  & 662$^{+48}_{-96}$                 \\
\vspace{5pt}
F                       & 1.22$\pm$0.06                 & 3.08$^{+0.15}_{-0.16}$        & 4.67$^{+0.79}_{-0.74}$  & 1.56$^{+0.26}_{-0.25}$  & 1202$^{+139}_{-84}$         \\
\vspace{5pt}
I                       & 1.22$^{+0.11}_{-0.12}$        & 1.86$^{+0.16}_{-0.18}$        & 2.66$^{+1.05}_{-0.88}$  & 1.47$^{+0.58}_{-0.49}$  & 1107$^{+325}_{-146}$        \\
\vspace{5pt}
J                       & 2.63$^{+0.22}_{-0.28}$        & 1.48$^{+0.12}_{-0.16}$        & 1.78$^{+0.56}_{-0.38}$  & 2.66$^{+0.84}_{-0.57}$  & 480$^{+25}_{-24}$           \\
\hline
\end{tabular}
\tablefoot{
\tablefoottext{a}{$\eta$ is the volume filling factor.}
The columns are the region name, electron number density, total gas mass, total thermal energy, thermal energy density, and cooling time.
}
\end{table*}

\begin{table*}[ht]
\caption{\label{diskgas}Derived parameters of the hot gas in and above the spiral arm regions of NGC\,6946.}
\centering
\begin{tabular}{cccccc}
\hline\hline
Region & (n$_{disk}\eta^{-0.5}$)\tablefootmark{a} & (M$^{disk}_{gas}\eta^{0.5}$)\tablefootmark{a}& (E$^{disk}_{th}\eta^{0.5}$)\tablefootmark{a}& ($\epsilon^{disk}_{th}\eta^{-0.5}$)\tablefootmark{a} & ($\tau^{halo}\eta^{0.5}$)\tablefootmark{a} \\
\vspace{5pt}&[10$^{-3}$cm$^{-3}$]                 & [10$^6$M$_\odot$]                             & [10$^{54}$\,erg]                                       & [10$^{-12}$\,erg\,cm$^{-3}$]   &  [Myr]     \\
\hline
\vspace{5pt}
A$_{disk}$      & 4.13$^{+0.52}_{-0.40}$        & 1.14$^{+0.15}_{-0.11}$        & 2.39$^{+0.56}_{-0.41}$  & 7.25$^{+1.69}_{-1.24}$        &       517$^{+47}_{-34}$       \\
\vspace{5pt}
A$_{halo}$      & 1.69$^{+0.15}_{-0.19}$        & 4.68$^{+0.42}_{-0.54}$        & 3.75$^{+0.78}_{-0.91}$        & 1.14$^{+0.24}_{-0.28}$  & 579$^{+12}_{-10}$           \\
\vspace{5pt}
E$_{disk}$      & 2.92$^{+0.47}_{-0.49}$        & 1.05$^{+0.17}_{-0.18}$        & 2.41$^{+0.70}_{-0.76}$  & 5.62$^{+1.63}_{-1.77}$        &       815$^{+79}_{-83}$       \\
\vspace{5pt}
E$_{halo}$      & 1.26$^{+0.16}_{-0.24}$        & 4.53$^{+0.56}_{-0.87}$        & 3.89$^{+1.21}_{-1.06}$        & 0.91$^{+0.28}_{-0.25}$  & 809$^{+90}_{-22}$           \\
\vspace{5pt}
G$_{disk}$      & 4.63$^{+0.31}_{-0.55}$        & 2.15$^{+0.14}_{-0.26}$        & 4.79$^{+0.72}_{-0.90}$  & 8.70$^{+1.30}_{-1.62}$        &       498$^{+42}_{-18}$       \\
\vspace{5pt}
G$_{halo}$      & 2.15$\pm$0.01                 & 9.96$\pm$0.07                 & 7.70$^{+0.63}_{-0.62}$        & 1.40$\pm$0.11           & 448$^{+37}_{-31}$           \\
\vspace{5pt}
H$_{disk}$      & 3.65$^{+0.37}_{-0.49}$        & 3.19$^{+0.32}_{-0.42}$        & 6.21$^{+1.33}_{-1.22}$  & 5.95$^{+1.28}_{-1.17}$        &       524$^{+63}_{-8}$        \\
\vspace{5pt}
H$_{halo}$      & 1.31$^{+0.14}_{-0.16}$        & 11.44$^{+1.22}_{-1.43}$       & 9.18$^{+1.34}_{-2.01}$        & 0.88$^{+0.13}_{-0.19}$  & 752$^{+39}_{-83}$           \\
\hline
\end{tabular}
\tablefoot{
\tablefoottext{a}{$\eta$ is the volume filling factor.}
The columns are the region name, electron number density, total gas mass, total thermal energy, thermal energy density, and cooling time.
}
\end{table*}

\subsubsection{Magnetic fields in NGC\,6946}
\label{bfields}

To analyse the magnetic field parameters we used the same regions as for the spectral analysis.
For each region both total intensity and polarized intensity fluxes at a radio wavelength of 6.2\,cm
with a beam of 15$\arcsec$ were obtained.
Then, using the energy equipartition formula provided by \citet{beckra}, we calculated the strengths of both the total and ordered magnetic fields. The calculations
were made assuming a synchrotron spectral index of 1.0, an inclination of the galactic disk of 30$\degr$, and a proton-to-electron ratio of 100.
For the emitting volume a disk of 1\,kpc thickness was assumed. The main uncertainties are introduced by the last two parameters. They may vary by a factor of 2. With the assumed
spectral index this amounts to an error of $\sim$\,30\% for the strength of the magnetic field and $\sim$\,60\% for its energy density. We note here, however, that such errors are
systematic, which means that we may in fact expect the relative uncertainties between the points to be smaller.
Table~\ref{magparams} summarizes our results and also provides values for the energy densities of the magnetic field.

\begin{table*}[ht]
\caption{\label{magparams} Properties of the magnetic fields in NGC\,6946.}
\centering
\begin{tabular}{cccccc}
\hline\hline
Region          &S$_{synch}$    &p$_{synch}$    &B$_{tot}$      &$\epsilon_B$                   &B$_{ord}$\\
\vspace{5pt}    &[mJy/beam]     &[\%]           &[$\mu$G]       &[10$^{-12}$\,erg\,cm$^{-3}$]   &[$\mu$G]\\
\hline
\vspace{5pt}
A               &  0.67         &   7.7         &  17.2$\pm$5.2         &   11.8$\pm$7.1                        &  4.9$\pm$1.5     \\
\vspace{5pt}
B               &  0.36         &  22.6         &  14.4$\pm$4.3         &    8.3$\pm$5.0                        &  7.2$\pm$2.2     \\
\vspace{5pt}
C               &  0.31         &  26.3         &  13.8$\pm$4.1         &    7.6$\pm$4.6                        &  7.4$\pm$2.2     \\
\vspace{5pt}
D               &  1.74         &   9.3         &  21.8$\pm$6.5         &   18.9$\pm$11.3                       &  6.8$\pm$2.0     \\
\vspace{5pt}
E               &  0.42         &  12.7         &  15.2$\pm$4.6         &    9.1$\pm$5.5                        &  5.6$\pm$1.7     \\
\vspace{5pt}
F               &  0.39         &  16.5         &  14.8$\pm$4.4         &    8.8$\pm$5.3                        &  6.2$\pm$1.9     \\
\vspace{5pt}
G               &  0.54         &   6.5         &  16.3$\pm$4.9         &   10.6$\pm$6.4                        &  4.2$\pm$1.3     \\
\vspace{5pt}
H               &  0.39         &  10.8         &  14.9$\pm$4.5         &    8.9$\pm$5.3                        &  5.0$\pm$1.5     \\
\vspace{5pt}
I               &  0.35         &  20.5         &  14.3$\pm$4.3         &    8.2$\pm$4.9                        &  6.8$\pm$2.0     \\
\vspace{5pt}
J               &  1.40         &  14.6         &  20.4$\pm$6.1         &   16.6$\pm$10.0                       &  8.1$\pm$2.4     \\
\hline
\end{tabular}
\tablefoot{
The columns are the region name, non-thermal radio flux, degree of polarization, total magnetic field strength, magnetic field energy, and ordered magnetic field strength.
}
\end{table*}

Apart from the central regions of the galaxy (regions D and J), the strength of the total magnetic field is roughly constant across the disk. A slight increase
can be observed in the most prominent spiral arms (regions A and G). Consequently, these regions show higher energy densities of the magnetic field, with the maximum
near the galactic core. The strengths of the ordered magnetic field are also similar in all parts of the disk, with higher values in the areas
of the magnetic arms and the central region of the galaxy. As suggested by Fig.~\ref{6946radio}, the areas of the magnetic arms show
a much higher degree of polarization than the other regions.

\subsubsection{Hot gas and magnetic fields of spiral and magnetic arms}
\label{spirals}

 In addition to the grand-design structure of its gaseous spiral arms, NGC\,6946 presents a distinct spiral structure of the magnetic fields which resembles
magnetic arms that are phase-shifted with regard to the gaseous ones. These magnetic arms coincide well with the interarm regions. Since the spiral arm and interarm regions vary
significantly in terms of the ISM structure, we investigated the emission from the hot gas in both areas to obtain more clues about the interplay of the magnetic field and the hot
plasma. Because a model with two thermal components was used to analyse the emission from the spiral arms (accounting for the disk and halo components), to compare it with the emission
from the hot gas in and above the interarm (magnetic arm) areas, we needed to calculate averages of the values obtained from the two-temperature fits. 

We added gas masses and thermal energies for the appropriate regions. Next, we calculated number and energy densities, 
taking into account the volumes assumed for the disk and halo component emission (the area of a given region times 1 or 10\,kpc, respectively). 
Our results are presented in Table~\ref{averagearms}. 
We compared the averaged values of number and energy densities for all spiral arm regions  with respective values for 
the interarm regions. We also calculated the ratios of the thermal energy densities to number densities, hence 
obtaining an average energy per particle, which is
independent of the unknown volume-filling factor ($E_p = \epsilon/n$).

\begin{table*}[ht]
\caption{\label{averagearms}Averaged parameters of the hot gas in and above the spiral arm regions of NGC\,6946.}
\centering
\begin{tabular}{ccccc}
\hline\hline
Region & ($n\,\eta^{-0.5}$)\tablefootmark{a} & (M$_{gas}\eta^{0.5}$)\tablefootmark{a}& (E$_{th}\eta^{0.5}$)\tablefootmark{a}& ($\epsilon_{th}\eta^{-0.5}$)\tablefootmark{a} \\
\vspace{5pt}&[10$^{-3}$cm$^{-3}$]                 & [10$^6$M$_\odot$]                             & [10$^{54}$\,erg]                                       & [10$^{-12}$\,erg\,cm$^{-3}$]       \\
\hline
\vspace{5pt}
A               & 1.91$^{+0.19}_{-0.21}$        & 5.82$^{+0.57}_{-0.66}$        & 6.14$^{+1.34}_{-1.32}$        & 1.70$^{+0.37}_{-0.38}$             \\
\vspace{5pt}
E               & 1.41$^{+0.18}_{-0.27}$        & 5.58$^{+0.72}_{-1.05}$        & 6.30$^{+1.91}_{-1.82}$        & 1.34$^{+0.40}_{-0.39}$             \\
\vspace{5pt}
G               & 2.38$^{+0.03}_{-0.06}$        & 12.11$^{+0.21}_{-0.33}$        & 12.49$^{+1.35}_{-1.51}$        & 2.06$^{+0.22}_{-0.25}$           \\
\vspace{5pt}
H               & 1.52$^{+0.16}_{-0.19}$        & 14.63$^{+1.54}_{-1.85}$        & 15.39$^{+2.65}_{-3.22}$        & 1.35$^{+0.23}_{-0.29}$           \\
\hline
\end{tabular}
\tablefoot{
\tablefoottext{a}{$\eta$ is the volume filling factor.}
The columns are the region name, electron number density, total gas mass, total thermal energy, and thermal energy density.
}
\end{table*}

We present our results in Table~\ref{particles}. The interarm regions show higher values of an energy per particle than for the spiral arm regions. 
This is consistent with the single-temperature fits (Fig.~\ref{singles}), which show slightly higher temperatures for the regions of the magnetic arms. 
This means that there may be an additional effect that provides thermal energy to the interarm regions. Since this additional heating might also be due to magnetic reconnection, 
we analysed the magnetic field properties of the areas of the spiral arms and the interarm regions.

\begin{table*}[ht]
\caption{\label{particles}Thermal energy per particle and energy densities of the magnetic field for regions of the spiral and magnetic arms of NGC\,6946.}
\centering
\begin{tabular}{ccrccr}
\hline\hline
Spiral arm      & $E_p$                 & $\epsilon_{B}$& Magnetic arm& $E_p$                   & $\epsilon_{B}$        \\
\hline
\vspace{5pt}
A               & 0.89$^{+0.10}_{-0.11}$& 11.8$\pm$7.1   & B           &1.45$^{+0.14}_{-0.15}$  & 8.3$\pm$5.0           \\
\vspace{5pt}
E               & 0.95$^{+0.14}_{-0.12}$& 9.1$\pm$5.5    & C           &1.33$^{+0.23}_{-0.27}$  & 7.6$\pm$4.6           \\
\vspace{5pt}
G               & 0.87$^{+0.08}_{-0.09}$& 10.6$\pm$6.4   & F           &1.28$^{+0.15}_{-0.16}$  & 8.8$\pm$5.3           \\
\vspace{5pt}
H               & 0.88$^{+0.06}_{-0.08}$& 8.9$\pm$5.3    & I           &1.20$^{+0.34}_{-0.31}$  & 8.2$\pm$4.9           \\
\hline
\end{tabular}
\tablefoot{
$E_p$ - energy per particle in 10$^{-9}$\,erg; $\epsilon_{B}$ - magnetic energy density in 10$^{-12}$\,erg\,cm$^{-3}$
}
\end{table*}

\subsubsection{Heating of the gas by magnetic reconnection?}
\label{reconn}

Because for the interarm regions we have the information about the hot gas coming from both the disk and the above halo 
(one-temperature fit to the spectra), a direct comparison of the energy densities of disk hot gas and magnetic fields is possible for the regions of the spiral arms alone.
Nevertheless, since we are interested in the global energy budget of the galaxy, we need information about both thermal and magnetic energy densities for the disk and the halo.
Although halo magnetic fields surely exist (as observed in edge-on spiral galaxies), we do not have any direct information
on their structure and strength in the case of NGC\,6946. The observed radio emission is, however, an integration along the line of sight, that is, we see
contributions from both the galaxy disk and halo. Since the majority of the cosmic rays originates in the underlying disk, an assumption for the emitting
volume (the disk) seems to be justified. As the vertical scale height of the halo magnetic fields of $\sim6-7$\,kpc \citep[e.g.][]{beck15} is similar to the 
assumed size of the hot gas halo (10\,kpc), we do not expect a significant change of the magnetic energy density in the halo, especially when an uncertainty of its 
calculation (60\%) is considered. It is therefore justified to compare the obtained magnetic energy densities with those of the hot gas in and above the interarm regions.

Still, for all areas of the disk of NGC\,6946 we see much higher energy densities of the magnetic fields
than those of the disk component of the hot gas (Tables~\ref{diskgas} and~\ref{magparams}). Interestingly, for the region of the most prominent spiral arm (region G),
conditions closest to equilibrium are observed, with the energy density of the magnetic field only 22\% higher than that of the hot gas in the galactic disk. For the remaining spiral
arms this difference is as high as $50-63\%$. If we compare the magnetic field energy densities to those of the halo (Tables~\ref{halogas} and~\ref{diskgas}) or 
averaged values for the hot gas in and above the spiral arms (disk+halo, Table~\ref{averagearms}), a distinct dominance of magnetic fields by a factor of a few is visible. 
This suggests that only in the areas of high star-forming activity it is possible that the thermal
energy density of the gas is similar to that of the local magnetic fields.

\begin{figure}[ht]
                        \resizebox{\hsize}{!}{\includegraphics[clip, angle=-90]{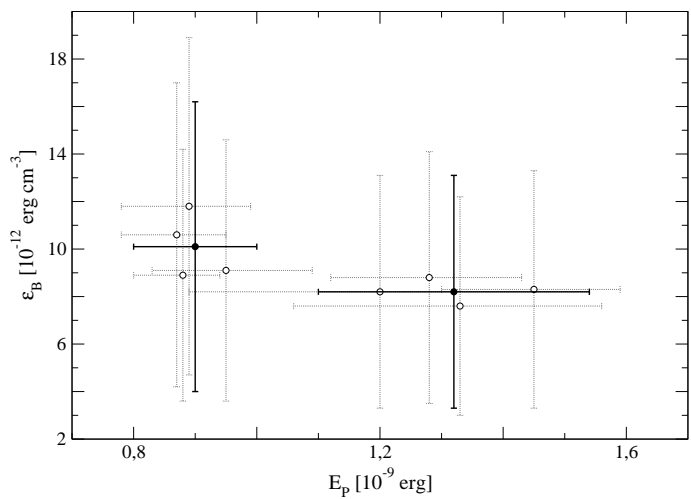}}
                \caption{Relation between energy per particle (${\rm E}_{\rm P}$) and energy density of the magnetic field (${\rm\epsilon}_{\rm B}$) 
                        in the spiral and the magnetic arms of NGC\,6946.
                        Open symbols with dashed error bars present each spiral arm (left side of the plot) and magnetic arm (right side of the plot). 
                        Filled symbols with solid error bars present average values of the respective data points.}
                \label{epemag}
\end{figure}

To investigate  the interplay between the thermal gas and the magnetic fields in greater detail, we compared the averaged (i.e. from both disk and halo components) 
thermal energies per particle with the magnetic field energy densities for the spiral and magnetic arms. We calculated the average
values for the gaseous spiral arms and the magnetic arms, which resulted in an energy density of the magnetic field of about 10.1\,$\times\,10^{-12}$\,erg\,cm$^{-3}$ for the spiral arms and
8.2\,$\times\,10^{-12}$\,erg\,cm$^{-3}$ for the interarm regions and for energies per particle of 0.90$\times$\,10$^{-9}$\,erg and 1.32$\times$\,10$^{-9}$\,erg, respectively.
Our results are presented in Table~\ref{particles}. An anti-correlation between the magnetic field strength (and its energy) and the thermal energy of the gas is 
visible (Fig.~\ref{epemag}); with a slight decrease of the energy density of the magnetic field (by 23\%), the energy per particle increases significantly (by 68\%). 

A possible explanation is that in regions that simultaneously
show higher thermal energies of the gas and lower energies of the magnetic fields, 
some of the energy of the magnetic field might have been converted into thermal energy by magnetic reconnection. 
Fast reconnection should be possible in most astrophysical plasmas \citep{hanasz}, with a heating rate proportional to the Alfv\'en speed
\citep{lesch,lazarian}. As the gas density in the interarm regions is lower while the total magnetic field is almost as strong as in the
spiral arms, the Alfv\'en speed is higher, and hence the heating rate is higher in interarm regions.
Indeed, we do observe such an additional heating of the gas in the magnetic arm regions  (see Tables~\ref{singles} and~\ref{6946xtabr}). 

\begin{figure}[ht]
                        \resizebox{\hsize}{!}{\includegraphics[clip, angle=-90]{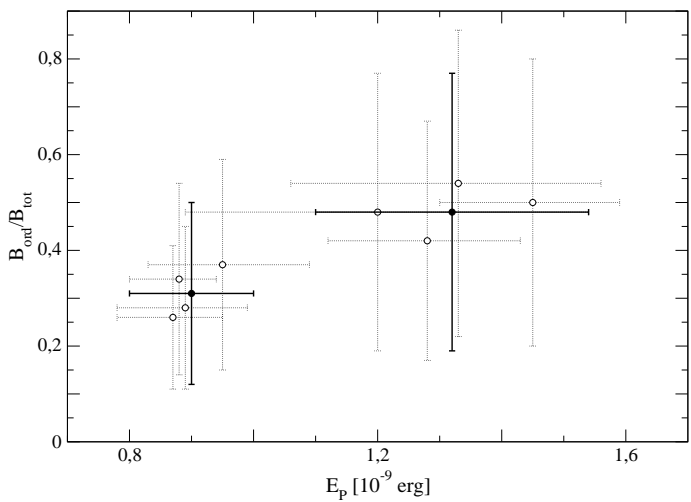}}
                \caption{Relation between energy per particle (${\rm E}_{\rm P}$) and regularity (${\rm B}_{\rm ord}/{\rm B}_{\rm tot}$) of the magnetic field 
                        in the spiral and the magnetic arms of NGC\,6946. Symbols are the same as used in Fig.~\ref{epemag}.}
                \label{epregul}
\end{figure}

In general, such a slight increase in temperature in the interarm regions could be easily explained 
by longer cooling times that are due to the lower density of the gas. However, if we compare interarm regions B and C with F and I, we note that in magnetic arms B and C 
higher temperatures, lower number and energy densities than in magnetic arms F and I were obtained. Surprisingly, 
the difference in cooling times reaches almost a factor of 2. 
Still, for all four interarm regions practically the same magnetic field strengths and energy densities were observed. 
The difference is visible, however, when ordered fields are considered -- in both regions B and C we see 
a much higher regularity of the field (i.e. the ratio of the strengths of ordered and total magnetic field). This trend is visible in all magnetic 
arms when compared to the spiral arms -- the regularity of the magnetic field increases with the energy per particle (Fig.~\ref{epregul}).

The above findings allow constructing a picture of both turbulence and reconnection acting in a galactic disk. Although in the spiral arms 
reconnection effects are expected to be more efficient (stronger field tangling), their action might be difficult to see because both the heating and field 
(dis)ordering is dominated by turbulence. 
In the interarm regions, however, where the magnetic field is highly ordered, reconnection heating may dominate turbulence heating, 
and the increase in temperature due to reconnection heating might be noticeable. This is what we observe, 
especially in the magnetic arms B and C, 
which have slightly higher temperatures than most star-forming 
spiral arms. Although reconnection is a very local process, acting at distances of a few pc or less, if it is equally efficient throughout the entire magnetic arm, 
it might contribute to the field ordering, an effect that would last longer because of the weaker turbulence. Again, this high field ordering is most distinctly seen in magnetic arms B and C.

The region of the western spiral arm (region H) is interesting. While for the other spiral arm regions we observe a significant contribution
from the halo component (see Table~\ref{6946xfr}), region H shows almost equal contributions from the halo and disk components. Furthermore, the difference
of temperatures of the two components is the lowest in this region, owing mainly to the lowest temperature of all spiral arm disk components. This may
suggest that for the western spiral arm mixing of the disk and the halo gas is the most efficient. Since the southern part of region H is in the area
of strong Faraday depolarization, as reported by \citet{beck91,beck07}, this might be a piece of evidence for vertical magnetic fields 
and enhanced outflow speed in this area of the galactic disk.
In particular the region of the south-western spiral arm (region A), which also contributes to the depolarization
area, shows a similar trend for the interplay of the halo and disk gas components, although at a lower level.

\section{Summary and conclusions}
\label{cons}

The detailed analysis of the X-ray emission from the hot gas in NGC\,6946, together with earlier radio continuum studies, can be summarized as follows:

\begin{itemize}
\item[-] A spectral analysis of the point sources revealed a significant number of ULXs, which agrees with the enhanced star formation of the galaxy.
\item[-] The galaxy presents significant emission from the hot gas across its entire star-forming disk. Intensity enhancements
are found in the regions of the spiral arms, also in the harder energy band. A significant amount of the very soft emission is found in the region of high Faraday depolarization.
\item[-] The radio-polarized emission structure is reflected in the spectral properties of the hot gas - areas of magnetic arms
visible in the interarm regions are well described with a single thermal plasma model, which shows that the temperature of the hot gas is
slightly higher than in the spiral arm regions.
\item[-] An increase in temperature of the hot gas in the magnetic arm regions could be described as additional heating
due to magnetic reconnection.
\item[-] A possible conversion of magnetic field energy into thermal energy of the hot gas in the interarm regions
is suggested by the lower energy density and strength of the magnetic field and the higher thermal energy per particle, when compared to the areas of
the spiral arms. 
\item[-] In the conditions of low turbulence in the magnetic arm regions, reconnection, acting mostly on tangled fields, might also contribute to 
the field ordering, as suggested by both the highest temperatures of the hot gas and the highest degree of polarization in magnetic arms B and C.
\item[-] We found signatures of a very hot gas in the area of the ultra-luminous source MF\,16, which may suggest shock heating
of the gas by a supernova explosion.
\end{itemize}

\begin{acknowledgements}
We thank Wolfgang Pietsch and Stefania Carpano for their collaboration on
the original XMM-Newton observing proposals that form the base of this paper.
Special thanks go to Harald Lesch and Alex Lazarian, who improved our
understanding of the reconnection theory. We also thank Stefanie
Komossa for useful comments on an earlier version of the paper, and
the anonymous referee for a helpful report.
\end{acknowledgements}

\bibliographystyle{aa} 
\bibliography{myreferences} 

\end{document}